\documentclass[11pt, a4paper, logo, copyright]{googledeepmind}

\pdftrailerid{redacted}

\makeatletter
\renewcommand\bibentry[1]{\nocite{#1}{\frenchspacing\@nameuse{BR@r@#1\@extra@b@citeb}}}
\makeatother

\usepackage{kantlipsum}
\usepackage{dsfont}
\usepackage{gdm-colors}
\usepackage[utf8]{inputenc}
\usepackage{newunicodechar}
\usepackage{wasysym,marvosym}
\usepackage[normalem]{ulem}
\usepackage{hyperref}
\usepackage{algorithmicx}
\usepackage{algpseudocode}
\usepackage{multirow}
\usepackage{geometry}
\usepackage[rightcaption]{sidecap}

\usepackage{tikz}
\usetikzlibrary{shapes.geometric, arrows.meta, positioning}
\usepackage{amsmath}
\usepackage{algorithm}
\usepackage{algpseudocode}
\usepackage{listings}
\usepackage{enumitem}
\usepackage{lipsum}
\usepackage[most]{tcolorbox}
\definecolor{thinkcolor}{RGB}{227,196,144}
\definecolor{observecolor}{RGB}{153,201,227}
\definecolor{explorecolor}{RGB}{178,217,200}

\newcounter{caseexample}[section]
\renewcommand{\thecaseexample}{\arabic{caseexample}}

\newcounter{promptexample}[section]

\newcommand{\assignmentQuestionName}{Question}

\usepackage{booktabs}
\usepackage{arydshln}
\usepackage{dashrule}
\usepackage{twemojis}

\usepackage[authoryear, sort&compress, round]{natbib}

\usepackage{bbding}
\usepackage[T1]{fontenc}
\usepackage{url}
\usepackage{nicefrac}
\usepackage{microtype}
\usepackage{amsmath}
\usepackage{graphicx}
\usepackage{multicol}
\usepackage[nameinlink]{cleveref}
\usepackage{bbm}
\usepackage{multirow}
\usepackage{soul}
\usepackage{float}
\usepackage{wrapfig}
\usepackage{blindtext}
\usepackage{tablefootnote}
\usepackage{amsfonts}
\usepackage[flushleft]{threeparttable}
\usepackage{colortbl}
\usepackage{mathtools}
\usepackage{bm}
\usepackage{CJKutf8}
\usepackage{makecell}
\usepackage{caption}
\usepackage{capt-of}
\usepackage{array}
\usepackage{calc}
\usepackage{caption}
\usepackage{subcaption}
\usepackage[bottom]{footmisc}
\usepackage{fontawesome}
\usepackage{tabularx}
\usepackage{siunitx}

\newcommand{\recgpt}[1][]{\textbf{\texttt{RecGPT#1}}}

\definecolor{hlcolor}{RGB}{206,32,74}
\newcommand{\hlnum}[1]{\textbf{\textcolor{hlcolor}{#1}}}
\newcommand{\prop}[1]{\textbf{\emph{#1}}}
\newcommand{\oldway}[1]{\textbf{\textcolor{teal!65!black}{#1}}}
\newcommand{\newway}[1]{\textbf{\textcolor{orange!85!black}{#1}}}
\newcommand{\metric}[1]{\textsf{\bfseries #1}}

\newtcbox{\pill}[1][observecolor]{on line, nobeforeafter,
  colback=#1!16, colframe=#1!16, boxrule=0pt, arc=4.5pt,
  left=4pt, right=4pt, top=1.2pt, bottom=1.2pt,
  fontupper=\small\ttfamily\bfseries, coltext=#1!60!black}
\newcommand{\sidp}{\pill[hlcolor]{sid}}
\newcommand{\txtp}[1]{\pill{#1}}

\newcolumntype{C}{>{\centering\arraybackslash}X}
\newcolumntype{L}{>{\raggedright\arraybackslash}X}

\title{RecGPT-V3 Technical Report}

\author{RecGPT Team}

\begin{abstract}
Large language models (LLMs) are transforming recommender systems from matching co-occurrence patterns in historical behavior toward reasoning about the intent that drives it. \recgpt[-V1] pioneered this paradigm on Taobao by placing a genuine understanding of the user at the center of the pipeline, and \recgpt[-V2] scaled it through coordinated multi-agent reasoning; both have been deployed in production and yielded consistent gains in user experience and commercial interests. Yet operating the \recgpt{} series at scale exposes three fundamental challenges: (1) \emph{stateless behavior modeling}, where each request reprocesses a user's full history from scratch, incurring redundant computation and discarding prior analysis; (2) a \emph{tag-to-item information bottleneck}, where natural-language tags form a lossy information channel between user understanding and item grounding; and (3) \emph{inefficient explicit reasoning}, whose lengthy chain-of-thought imposes untenable latency and computational overhead.

To address these challenges, we present \recgpt[-V3], a stateful, hybrid-modal recommender that efficiently reasons over natural language for open-world knowledge and Semantic IDs (SIDs) for concrete item grounding. 
As the system engine, a \emph{Memory Hub} maintains a structured, continually evolving user memory that distills long-horizon behavior into condensed memory units, reducing user-modeling computation by 55.8\%. 
To bridge reasoning and retrieval, a \emph{Hybrid-modal Foundation Model} enables the LLM to reason jointly over text-based tags and SIDs, opening a high-bandwidth channel into the item space. 
To render reasoning tractable under latency and compute budgets, \emph{Latent Intent Reasoning} internalizes verbose rationales into compact learnable latent tokens that remain decodable into readable explanations, lowering output token cost by $\bf{200\times}$. 
Deployed in Taobao's ``Guess What You Like'' feed, \recgpt[-V3] achieves consistent gains in large-scale online A/B tests, improving IPV by +1.28\%, CTR by +1.00\%, TC by +1.97\%, and GMV by +3.97\% while substantially cutting end-to-end serving resource consumption by 52.4\%. 
These results show that LLMs can serve as the intelligent brain of industrial-scale business systems, advancing recommendation accuracy and serving efficiency in tandem.
\end{abstract}

\begin{document}
\begin{CJK*}{UTF8}{gkai}

\maketitle

\vspace{-0.8em}
\begin{center}
\includegraphics[width=\linewidth]{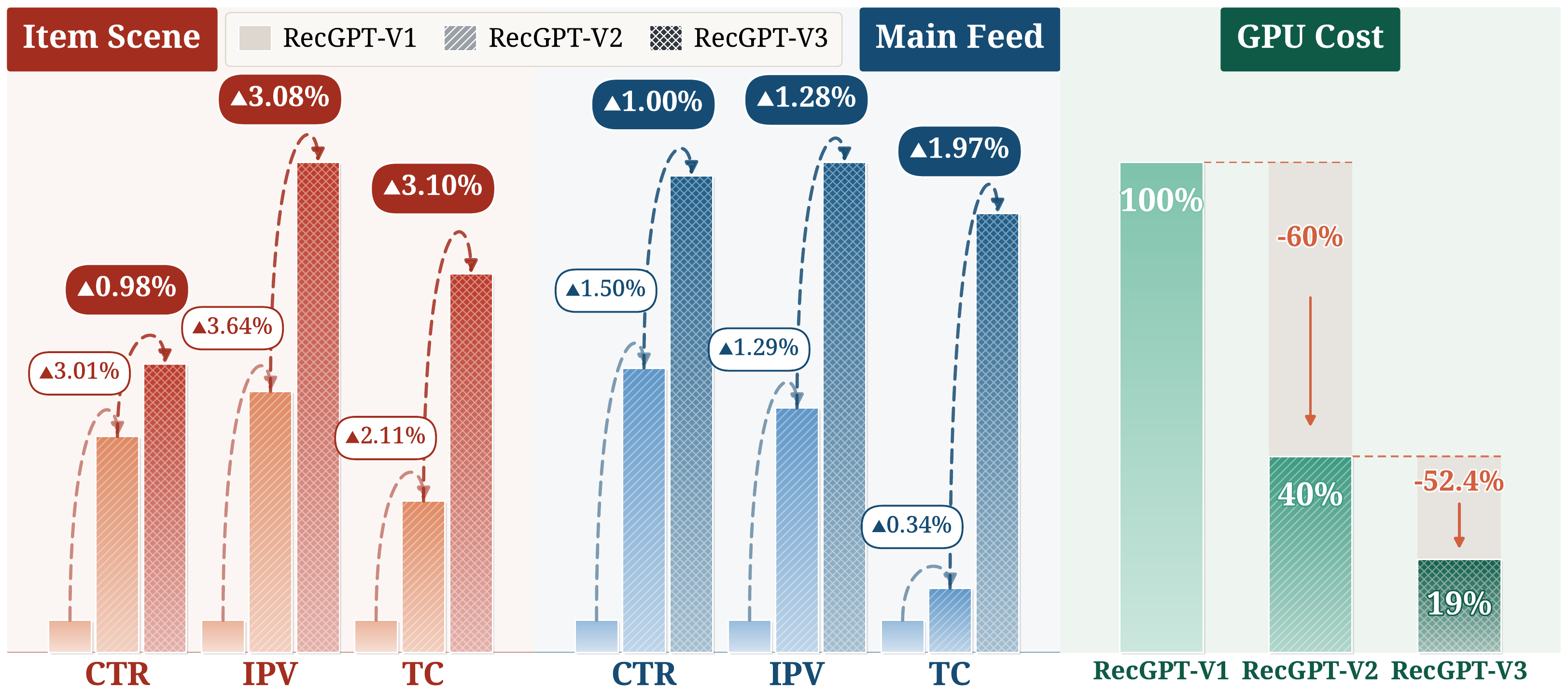}
\vspace{-1.5em}
\captionof{figure}{Performance and efficiency of the \recgpt{} series on Taobao. Each generation brings further gains on CTR, IPV, TC, and GMV in both the Item Scene and Main Feed, while GPU cost drops steadily: \recgpt[-V3] uses only 19\% of \recgpt[-V1]'s compute, a 52.4\% reduction over \recgpt[-V2].}
\label{fig:teaser}
\end{center}
\vfill

\newpage
\setcounter{tocdepth}{2}

\tableofcontents

\newpage

\section{Introduction}

An ideal recommender system should grasp what a user actually wants and surface the items that meet that underlying need.
In practice, however, the models that power today's platforms largely reduce this goal to predicting the next interaction from statistical regularities in historical logs.
From matrix factorization~\citep{rendle2012bpr,koren2009matrix} to deep sequential networks~\citep{kang2018self,tang2026think}, these models have grown increasingly expressive, yet they share a common limitation: they match co-occurrence patterns in past behavior, capturing which items appear together but not the intent behind them.
Confined to such correlational signals, they amplify what a user has already revealed, narrowing exposure~\citep{chen2023bias}, reinforcing filter bubbles~\citep{nguyen2014exploring}, and underserving long-tail content~\citep{gao2023alleviating}, which gradually erodes user experience and the health of the recommendation ecosystem.

The emergence of large language models (LLMs) opens a promising path beyond this limitation.
By drawing on broad world knowledge and deliberative reasoning~\citep{zhao2023survey,guo2025deepseek}, LLMs can interpret user behavior in context and infer the motivation and potential interest behind each action, rather than extrapolating co-occurrence alone.
This capability recasts recommendation as a problem of reasoning about user intent, a direction we have pursued across our work.
\recgpt[-V1]~\citep{yi2025recgptv1} rebuilt the industrial pipeline around this principle, delegating interest mining, item retrieval, and explanation generation to the LLM so that recommendations follow from understanding why a user acts, not from fitting what they did before.
Building on this foundation, \recgpt[-V2]~\citep{yi2025recgptv2} coordinated a planner and specialized multi-expert sub-agents to analyze user intent collaboratively, broadening its coverage while sharply lowering cost.
Both systems have been deployed at scale on Taobao, yielding consistent gains for users, merchants, and the platform and showing that LLM-driven user modeling is promising and practical in production.

Despite these gains, operating the \recgpt{} series at scale has exposed three key challenges that constrain further progress.
\begin{description}[leftmargin=2.4em, style=sameline, font=\normalfont, itemsep=4pt]
  \item[\colorbox{teal!85!black}{\textcolor{white}{\textbf{~C1~}}}] \textbf{Stateless behavior modeling.}
  Most existing LLM-based recommenders~\citep{zhou2025openonerec,team2026onereason,wang2026llm} reason over a user's entire behavior history in a one-shot manner.
  As interactions accumulate, each request reprocesses the full history from scratch, without reusing any earlier analysis.
  This stateless design incurs redundant computation that grows with the behavior sequence and, more fundamentally, prevents the system from carrying forward what it has already learned about the user.

  \item[\colorbox{teal!85!black}{\textcolor{white}{\textbf{~C2~}}}] \textbf{Tag-to-Item information bottleneck.}
  Both \recgpt[-V1] and \recgpt[-V2] have the LLM predict natural-language tags of the items a user is likely to interact with next, which a downstream model then uses to ground concrete items.
  This textual interface forms a lossy channel between reasoning and retrieval: a coarse tag (e.g., ``\textit{rotating adjustable badminton racket stand}'') corresponds to a broad and heterogeneous set of candidate items and cannot convey the fine-grained, ID-level evidence that the item space relies on, leaving a unbridgeable information gap between the intent the LLM predicts and the items ultimately retrieved.

  \item[\colorbox{teal!85!black}{\textcolor{white}{\textbf{~C3~}}}] \textbf{Inefficient explicit reasoning.}
  Explicit chain-of-thought helps the LLM understand users and uncover their latent interests~\citep{liu2025onerec,li2026gr2technicalreport,zhang2026reasonrecreasoningaugmentedmultimodalagent}, but autoregressively generating lengthy rationales (averaging $\sim3,000$ tokens in our task) incurs considerable inference latency.
  At billion-user scale and under high QPS, this cost is prohibitive, naturally raising a central question: how to cut latency while retaining the explainable reasoning ability that the \recgpt{} series has upheld as a core design philosophy.
\end{description}

\begin{figure}[t]
\centering
\includegraphics[width=\textwidth]{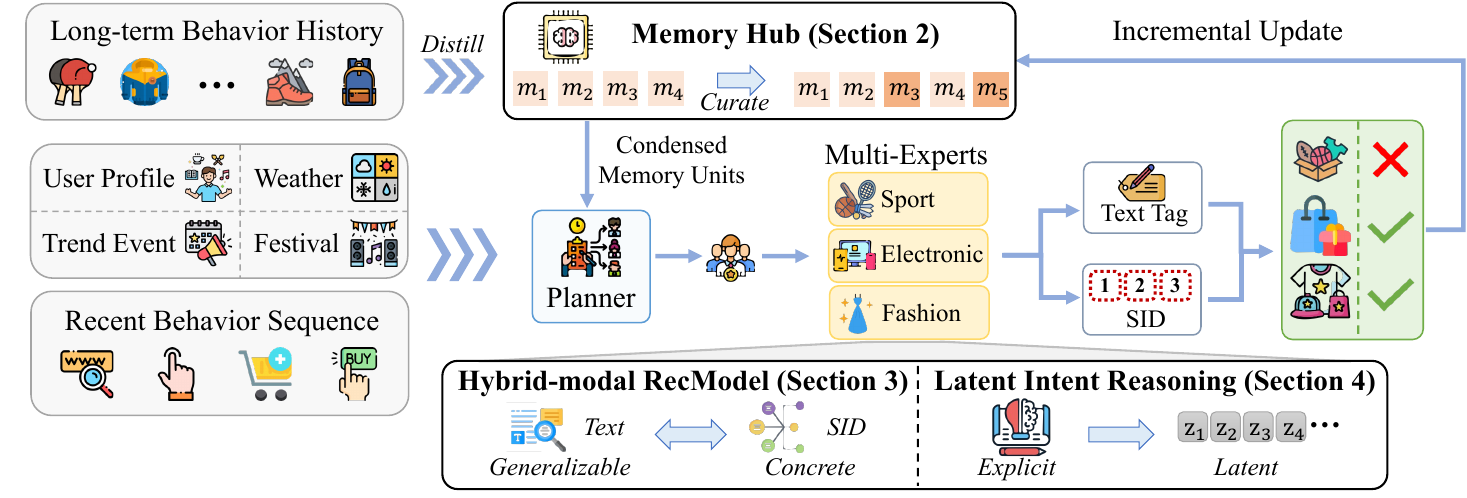}
\caption{Overview of \recgpt[-V3]. A \emph{Memory Hub} distills a user's long-term behavior into incrementally curated memory units; conditioned on these units and recent behaviors, a planner and its multi-expert modules reason with the \emph{Hybrid-modal Foundation Model} (natural language and Semantic IDs) and \emph{Latent Intent Reasoning} (compact latent tokens) to predict the next item to interact with.}
\label{fig:overview}
\end{figure}

To address these challenges, we improve \recgpt[-V3] along three dimensions, namely user modeling, intent representation, and intent reasoning, with the goal of jointly improving recommendation accuracy and efficiency.
Figure~\ref{fig:overview} illustrates the overall \recgpt[-V3] framework, which embodies the following three key technical innovations.
\begin{description}[leftmargin=2.4em, style=sameline, font=\normalfont, itemsep=4pt]
  \item[\colorbox{orange!85!black}{\textcolor{white}{\textbf{~S1~}}}] \textbf{Memory Hub}~(\S\ref{sec:memory}).
  We move from \oldway{stateless, one-shot user modeling} to a \newway{stateful, memory-augmented recommender} that continually learns each user's personalized preferences, placing a structured user memory at the center of the pipeline as its engine.
  This memory distills each user's long-horizon behavior history into succinct memory units, compressing tokens by \hlnum{94.5\%} (\prop{condensed}), links every unit back to its originating behaviors for provenance (\prop{traceable}), and curates them incrementally as new interactions arrive (\prop{evolvable}), so that understanding of the user accrues over time, no longer rediscovered from scratch.
  The model thus draws on the consolidated memory together with a recent behavioral increment instead of the raw full history, reducing user-modeling computation by \hlnum{55.8\%} while preserving essential behavioral contexts.

  \item[\colorbox{orange!85!black}{\textcolor{white}{\textbf{~S2~}}}] \textbf{Hybrid-modal Recommendation Foundation Model}~(\S\ref{sec:foundation}).
  To overcome the \oldway{Tag-to-Item} bottleneck, we equip the LLM to reason in \newway{\prop{Semantic IDs (SIDs)}}, discrete codes that encode item semantics, as a second modality alongside natural language.
  The two modalities serve complementary purposes: natural language expresses intent through rich, open-ended world knowledge (\prop{generalizable}), while SIDs ground that intent in semantic item representations enriched by collaborative signals (\prop{concrete}).
  The model therefore generates intent representations natively compatible with the downstream model, opening a high-bandwidth channel from intent reasoning to product retrieval that closes the gap left by the language-only interface.

  \item[\colorbox{orange!85!black}{\textcolor{white}{\textbf{~S3~}}}] \textbf{Latent Intent Reasoning}~(\S\ref{sec:reasoning}).
  We internalize \oldway{lengthy explicit chain-of-thought} into \newway{compact learnable latent-tokens}, collapsing a verbose rationale into a few dense steps of latent deliberation.
  We introduce dedicated latent \texttt{<CoT>} slots that encode an entire reasoning trace, cutting its token cost by \hlnum{200$\bf{\times}$} (\prop{efficient}).
  Crucially, these latent tokens decode back into a readable rationale on demand (\prop{explainable}), reconciling low-latency inference with the interpretability that production recommendation demands.
\end{description}

\recgpt[-V3] has been deployed on the ``Guess What You Like'' scenario of the Taobao homepage, one of the world's largest e-commerce platform surfaces serving hundreds of millions of daily active users.
Through large-scale online A/B testing against the live \recgpt[-V2] system, \recgpt[-V3] delivers strong gains across engagement and business metrics, including \hlnum{+1.28\%} IPV, \hlnum{+1.00\%} CTR, \hlnum{+1.97\%} TC, and \hlnum{+3.97\%} GMV, while simultaneously reducing end-to-end serving compute by \hlnum{52.4\%}.
More broadly, \recgpt[-V3] shows that large language models can power the real-world industrial pipeline at low cost, returning recommendation application to its founding purpose: to faithfully understand and serve what every user truly needs.

\section{Memory Hub}
\label{sec:memory}
In the RecGPT-V2 pipeline, the Global Planner analyzes a user's behavioral history, decomposes it into intent personas, and dispatches them to multi-expert modules for item tag prediction and retrieval. Since it reprocesses the full behavioral sequence at every pass (${\sim}$\hlnum{55K} tokens for highly active users), the Planner alone accounts for ${\sim}$\hlnum{95\%} of the computational cost in agentic intent analysis, while each pass redundantly rediscovers patterns such as repurchase cycles, brand loyalty, and seasonal shifts that could instead persist and update across sessions. To address this, we introduce the \textbf{Memory Hub}, a persistent, structured memory layer that turns the Planner's input from stateless full-sequence encoding into stateful memory-driven reasoning, achieving three properties simultaneously:
 
\begin{itemize}
    \item \textbf{Condensed}: consolidates the full behavioral sequence into a compact set of schema-defined memory units, achieving \hlnum{80\%} token reduction while retaining only coherent, high-confidence behavioral patterns.
    \item \textbf{Traceable}: preserves provenance links from each memory unit to its source behaviors, enabling interpretability while maintaining downstream recommendation quality.
    \item \textbf{Evolvable}: incrementally incorporates new behaviors through selective updates and new pattern extraction, without full re-encoding.
\end{itemize}
 
\begin{figure*}[t]
  \centering
  \includegraphics[width=\textwidth]{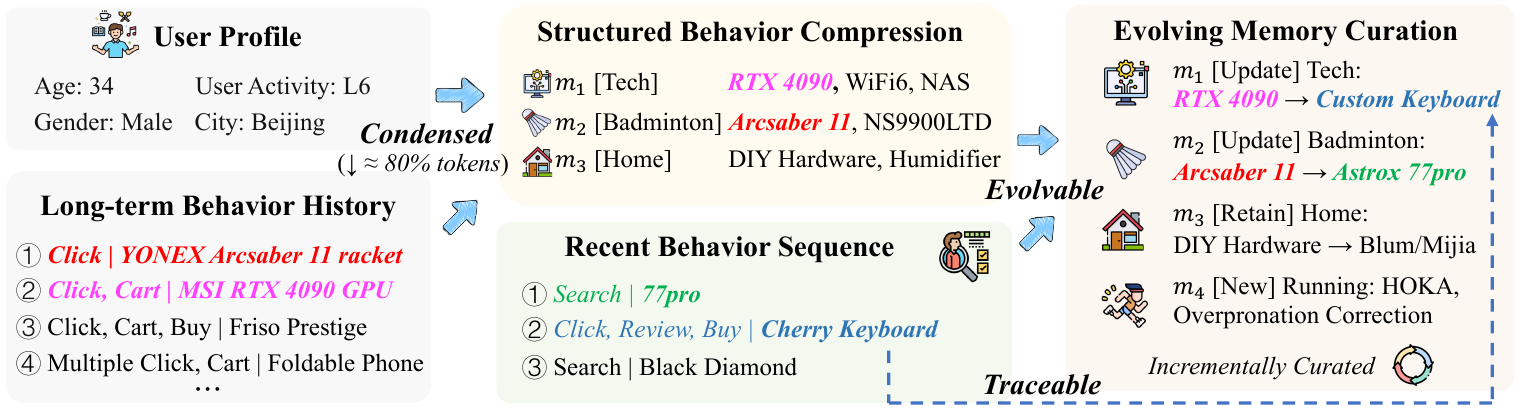}
  \caption{Overview of the Memory Hub. Structured Behavior Compression distills a user's long-term behavioral history into a compact set of structured memory units, reducing token usage by $94.5\%$. Evolving Memory Curation then keeps this memory current by selectively updating existing units and extracting new patterns from unmatched behaviors, yielding a condensed, traceable, and evolvable user representation for downstream recommendation.}
\label{fig:memory_hub_overview}
\end{figure*}

The memory hub builds and maintains this memory through two mechanisms: \textbf{Structured Behavior Compression} (\S\ref{sec:sbc}) performs an initial consolidation of each user's full historical behavior sequence into structured memory units, while \textbf{Evolving Memory Curation} (\S\ref{sec:emc}) keeps this memory current by periodically integrating newly observed behaviors. Together, they reduce overall Global Planner compute by \hlnum{55.8\%} while preserving recommendation quality. Figure~\ref{fig:memory_hub_overview} provides an overview of the memory hub architecture.

\subsection{Structured Behavior Compression}
\label{sec:sbc}

Structured behavior compression consolidates the user's full behavioral sequence $\mathcal{B} = \{b_1, b_2, \ldots, b_N\}$ into a compact set of structured memory units $\mathcal{M} = \{m_1, m_2, \ldots, m_K\}$ with $K \ll N$. Formally, we define this process as an LLM-based memory construction function:
\begin{equation}
    \mathcal{F}_{\phi}: \mathcal{B} \longrightarrow \mathcal{M}, 
    \qquad 
    \mathcal{M} = \mathcal{F}_{\phi}(\mathcal{B}) = \{m_1, m_2, \ldots, m_K\}, 
    \label{eq:sbc_map}
\end{equation}
It runs once as an initial pass, producing a persistent memory that serves as the foundation for subsequent incremental updates (\S\ref{sec:emc}). Each unit centers on two core fields that together characterize a behavioral pattern. The first, the \textbf{behavior pattern identifier} $p_k$, is a categorical label drawn from a predefined taxonomy of hundreds of patterns spanning the major e-commerce behavioral archetypes, with controlled extension permitted when no existing pattern captures a high-confidence cluster. The second, the \textbf{preference summary} $s_k$, describes in natural language the user's preferences, trends, and consumption characteristics within that pattern. Complementing these two fields, each unit further records supporting metadata: representative behavior indices for provenance, preferred brands, a temporal activity profile, and a timestamp. Table~\ref{tab:schema_example} shows a complete example.
 
\begin{table}[h]
\centering
\captionsetup{justification=centering}
\caption{Example of a structured memory unit.}
\label{tab:schema_example}
\small
\begin{tabularx}{\textwidth}{l X}
\toprule
\textbf{Field} & \textbf{Example} \\
\midrule
Behavior Pattern & K-pop Fandom \\
\addlinespace
Preference Summary & Dedicated fan of a K-pop girl group. Full-lifecycle engagement from album pre-orders to merchandise collection and live concert participation. Strong focus on collection completeness and item preservation. \\
\addlinespace
Representative Indices & 549, 553, 558, 563, 564, 565, \ldots \\
\addlinespace
Preferred Brands & SingBA, 时代良品, 珍琢, \ldots \\
\addlinespace
Temporal Activity & First appeared March 2023; peaked June and Aug--Oct 2023. \\
\addlinespace
Timestamp & Created: 2023-10-31 \\
\bottomrule
\end{tabularx}
\end{table}

\begin{table}[t]
\centering
\caption{Incremental curation of memory list from $\mathcal{M}^{(t)}$ to $\mathcal{M}^{(t+\delta)}$. Row colors indicate operation type: \colorbox{orange!10}{Update} (summary refreshed with new evidence), \colorbox{gray!6}{Retain} (unit left unchanged for lack of relevant new behaviors), and \colorbox{teal!10}{New} (unit created from behaviors unmatched by any existing unit).}
\label{tab:evolution}
\small
\begin{tabularx}{\textwidth}{c l X X l}
\toprule
\textbf{Unit} & \textbf{Pattern} ($p_k$) & \textbf{Summary at $t$} & \textbf{Summary at $t{+}\delta$} & \textbf{Operation} \\
\midrule
\rowcolor{orange!10}
$m_1$ & Infant Care & Frequent buyer of formula, diapers, and baby clothing for 0--6 months. $\cdots$ & $\cdots$ Shifted to \textbf{6--12 months}. Added \textbf{toddler toys and walkers}. & (1) Update \\
\addlinespace
\rowcolor{gray!6}
$m_2$ & Outdoor Running & Running shoes, GPS watches, and marathon nutrition supplements. $\cdots$ & (Unchanged) & (1) Retain \\
\addlinespace
\rowcolor{orange!10}
$m_3$ & Home Cooking & Budget-friendly kitchen gadgets and baking supplies. $\cdots$ & $\cdots$ Shifted toward \textbf{premium cookware and air fryer accessories}. & (1) Update \\
\addlinespace
\rowcolor{gray!6}
$m_4$ & Photography & Mirrorless camera accessories and lens filters. $\cdots$ & (Unchanged)  & (1) Retain \\
\addlinespace
\rowcolor{teal!10}
$m_5$ & \emph{Pet Parenting (new)} & --- & Pet food and grooming supplies. First-time pet owner. & (2) New \\
\bottomrule
\end{tabularx}
\end{table}

\subsection{Evolving Memory Curation}
\label{sec:emc}

Structured Behavior Compression (\S\ref{sec:sbc}) establishes the initial memory state, but user interests drift over time as preferences emerge, decay, and shift with life stages and seasons. A static memory would progressively diverge from the user and erode recommendation relevance. \textbf{Evolving Memory Curation} counters this drift through periodic incremental updates that fold newly observed behaviors into the existing memory. The Global Planner then reasons over this compact memory together with the recent behavioral delta rather than re-encoding the full behavioral sequence at each pass, thereby substantially reducing user-modeling computation.

Specifically, let $\mathcal{M}^{(t)} = \{m_1^{(t)}, \ldots, m_K^{(t)}\}$ denote the memory state at time $t$ produced by initial compression or a previous curation cycle, and let $\Delta\mathcal{B}^{(t, t+\delta)} = \{b_{N+1}, \ldots, b_{N+\Delta N}\}$ denote the newly accumulated behavioral interactions during the interval $[t, t+\delta)$. Evolving Memory Curation defines the incremental update function $\mathcal{G}$ as follows:
\begin{equation}
    \mathcal{G}:
    \left(\mathcal{M}^{(t)},\; \Delta\mathcal{B}^{(t,t+\delta)}\right)
    \longrightarrow
    \left(\mathcal{M}^{\mathrm{update}},\; \mathcal{M}^{\mathrm{new}}\right),
    \label{eq:incremental}
\end{equation}
where $\mathcal{M}^{\mathrm{update}}$ denotes the set of existing memory units after selective update or retention, and $\mathcal{M}^{\mathrm{new}}$ denotes the set of newly created memory units extracted from previously unmatched behaviors. The final curated memory is then obtained by merging these two parts. By design, $\mathcal{G}$ operates without access to the full sequence $\mathcal{B}$, drawing only on the compressed memory $\mathcal{M}^{(t)}$ and the new delta $\Delta\mathcal{B}^{(t,t+\delta)}$.

\paragraph{(1) Selective Update of Existing Units.}
For each existing memory unit $m_k^{(t)}$, the model determines whether the new behavioral delta contains interactions relevant to this pattern:
\begin{equation}
    m_k^{(t+\delta)} =
    \begin{cases}
        \text{Update}(m_k^{(t)},\; \Delta\mathcal{B}_k^{\mathrm{rel}}), 
        & \text{if } \Delta\mathcal{B}_k^{\mathrm{rel}} \neq \emptyset, \\
        m_k^{(t)}, 
        & \text{otherwise},
    \end{cases}
    \label{eq:selective_update}
\end{equation}
where $\Delta\mathcal{B}_k^{\mathrm{rel}} \subseteq \Delta\mathcal{B}^{(t,t+\delta)}$ denotes the subset of new behaviors semantically related to $m_k^{(t)}$, and $\Delta\mathcal{B}^{\mathrm{rel}}=\bigcup_{k=1}^{K} \Delta\mathcal{B}_k^{\mathrm{rel}}$ denotes the union of all new behaviors assigned to existing memory units. When triggered, the $\text{Update}$ operation refreshes the unit's summary $s_k$ and all supporting metadata; otherwise, the unit is retained without modification. This selective policy follows the principle of \prop{updating with evidence and retaining without disturbance}. The resulting updated-or-retained memory set is:
\begin{equation}
    \mathcal{M}^{\mathrm{update}}
    =
    \{\,m_k^{(t+\delta)}\,\}_{k=1}^{K}.
    \label{eq:updated_memory}
\end{equation}

\paragraph{(2) New Pattern Extraction.}
The interactions that the selective update leaves unassigned to any existing unit form the novel behavior set:
\begin{equation}
    \Delta\mathcal{B}^{\mathrm{novel}}
    =
    \Delta\mathcal{B}^{(t,t+\delta)}
    \setminus
    \Delta\mathcal{B}^{\mathrm{rel}}.
    \label{eq:novel}
\end{equation}
New pattern extraction is performed within the same curation function $\mathcal{G}$ in Eq.~\ref{eq:incremental}. Given the previous memory $\mathcal{M}^{(t)}$ and the new behavioral delta $\Delta\mathcal{B}^{(t,t+\delta)}$, $\mathcal{G}$ identifies coherent unmatched behaviors and writes them into new memory units:
\begin{equation}
    \mathcal{M}^{\mathrm{new}}
    =
    \text{Extract}\left(\Delta\mathcal{B}^{\mathrm{novel}}\right).
    \label{eq:new_memory}
\end{equation}
If no coherent new pattern is identified, $\mathcal{M}^{\mathrm{new}}=\emptyset$.

The final memory combines the existing units retained or updated by the selective update with these newly extracted ones:
\begin{equation}
    \mathcal{M}^{(t+\delta)}
    =
    \underbrace{\mathcal{M}^{\mathrm{update}}}_{\text{updated or retained}}
    \;\cup\;
    \underbrace{\mathcal{M}^{\mathrm{new}}}_{\text{newly extracted}}.
    \label{eq:merge}
\end{equation}
In practice, the two operations share a single model forward pass that jointly updates the existing units and extracts the new ones, sparing the pipeline from multiple sequential calls.

\paragraph{Semantic Continuity.}
Each selective update re-synthesizes the affected unit into a coherent snapshot of the user's \emph{current} state rather than appending new behaviors to the old summary, preventing stale and emerging preferences from accumulating within one unit. As Table~\ref{tab:evolution} shows, $m_1$ shifts from ``0--6 months'' to ``6--12 months'' and $m_3$ toward premium cookware as interests mature, while unaffected units ($m_2$, $m_4$) remain unchanged and a first-time interest forms the new unit $m_5$.
 
\label{sec:compute}
To summarize, the memory hub transforms each user's raw behavioral sequence into a structured, continually evolving memory that serves as the engine of the recommendation pipeline. Instead of re-deriving user understanding from the full interaction history at every request, the system reasons over this compact, high-signal memory together with a recent behavioral delta, so that its understanding of the user accrues and refines over time. In production, incremental curation runs every two months, reducing overall user-modeling computation by \hlnum{55.8\%}. Detailed efficiency analysis and quality validation are presented in \S\ref{sec:exp:memory}.
\section{Hybrid-Modal Recommendation Foundation Model}
\label{sec:foundation}
Text-only LLM experts interface with the discrete item space through a single natural-language channel: they predict free-form item tags that a downstream retriever grounds to in-domain items. This indirection makes the tag a \textbf{lossy bottleneck} between reasoning and retrieval: a coarse tag such as ``lightweight running shoes for marathon training'' matches a broad, heterogeneous set of items and cannot convey the fine-grained, item-level evidence that retrieval depends on, loosening the coupling between the intent the model expresses and the items ultimately surfaced.

\begin{figure}[t]
\centering
\includegraphics[width=\textwidth]{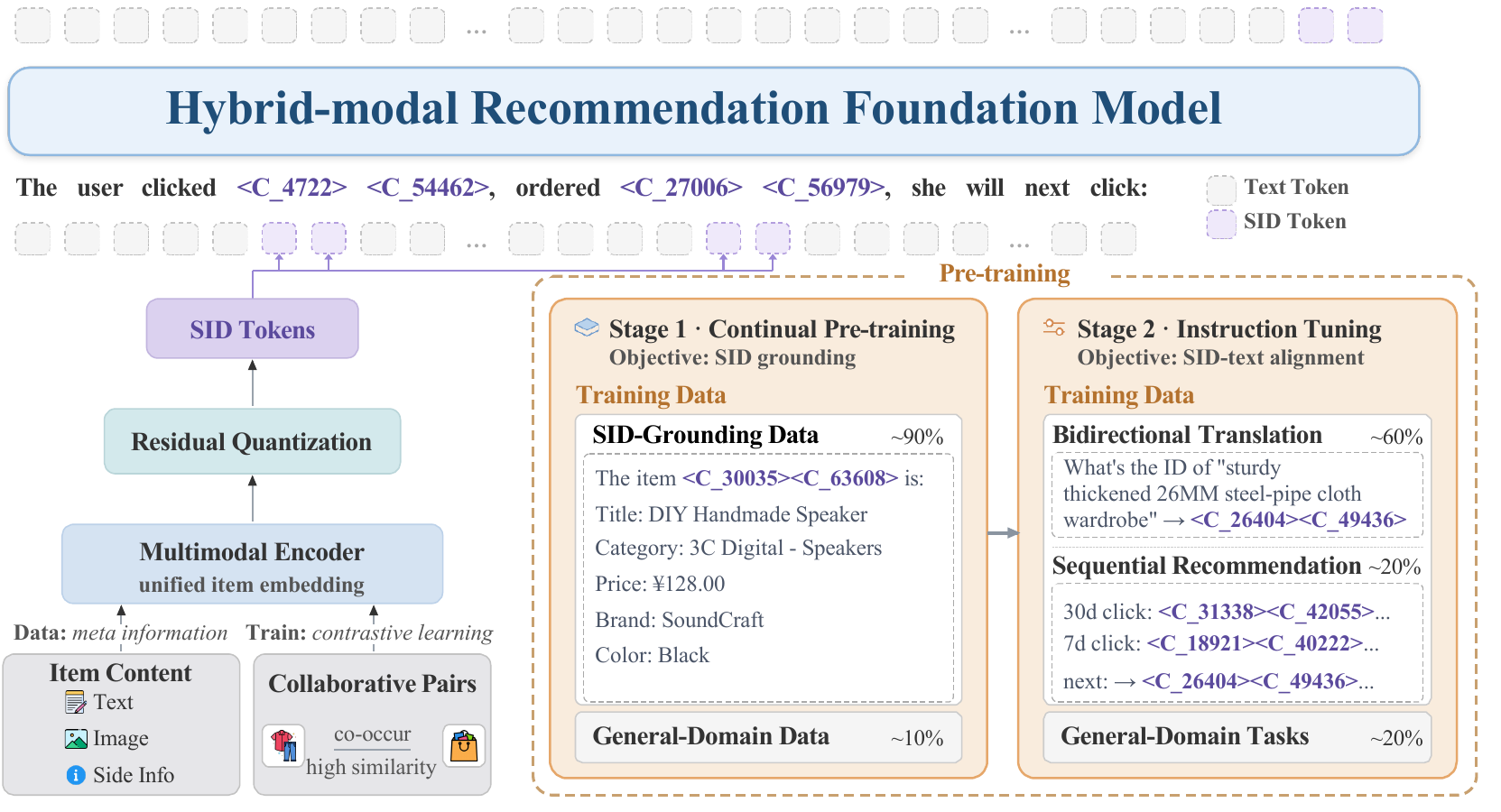}
\caption{Overview of the Hybrid-modal Foundation Model. Multimodal item features are quantized into $65{,}536$ SID tokens via CN-CLIP and a two-level RQ-VAE, extending the Qwen3-14B vocabulary. Continual pre-training (Stage~1) and instruction tuning (Stage~2) then align these tokens with language, with general-domain data mixed into both stages.}
\label{fig:foundation_overview}
\end{figure}

To overcome this bottleneck, we propose a \textbf{Hybrid-Modal Recommendation Foundation Model} that extends the Qwen3-14B~\citep{yang2025qwen3technicalreport} backbone with \prop{Semantic IDs (SIDs)}—discrete codes that encode item semantics—as a second modality alongside natural language. Recommendation inherently requires reasoning about user intent while simultaneously grounding it in concrete items, a dual demand that no single representation satisfies in isolation. The two modalities are thus complementary, each addressing one facet of this requirement:
\begin{itemize}
    \item \prop{Generalizable}: natural language expresses intent through rich, open-ended world knowledge.
    \item \prop{Concrete}: SIDs ground that intent in item representations enriched by collaborative signals.
\end{itemize}
By incorporating both modalities into a unified vocabulary, the model emits retrieval-compatible identifiers rather than coarse tags, thereby opening a high-bandwidth channel into the item space while preserving its full natural-language capability. Figure~\ref{fig:foundation_overview} illustrates the overall pipeline, comprising two components: a hybrid tokenization scheme that constructs semantically grounded SIDs (\S\ref{sec:hybrid_tokenization}), followed by a two-stage training procedure that grounds these tokens in language and equips the model to reason jointly over SIDs and text (\S\ref{sec:foundation:cpt}).

\subsection{Hybrid Tokenization}
\label{sec:hybrid_tokenization}

Hybrid tokenization augments the LLM's native text vocabulary with Semantic ID tokens derived from item content, so that semantically similar items receive similar codes. Constructing these SIDs follows our prior FORGE framework~\citep{fu2025forge}: we first learn a discriminative multimodal item representation via contrastive learning, then quantize it into discrete hierarchical codes via Residual Quantization (RQ-VAE)~\citep{Lee2022AutoregressiveIG,chen2026shopx}. We detail each below.

\paragraph{Multimodal Item Representation.}
The goal of this stage is to obtain a unified, discriminative embedding for each item by fusing its text, image, and side information. As no explicit labels define item similarity, we mine positive pairs directly from behavior: items that frequently co-occur across the behavioral logs form candidate pairs, from which we discard those whose multimodal similarity is low, guarding against spurious co-occurrences. The surviving pairs supervise contrastive learning.

Specifically, for each item $i$, we extract its text description $I^i_{\text{text}}$, image information $I^i_{\text{image}}$, and auxiliary side information $I^i_{\text{side}}$ (\textit{e.g.}, item attributes and metadata). We employ CN-CLIP~\citep{yang2023chineseclipcontrastivevisionlanguage}, denoted $\mathcal{E}$, to encode each modality, with the input determining which tower is invoked: $\mathcal{H}^i_{\text{text}} = \mathcal{E}(I^i_{\text{text}})$ and $\mathcal{H}^i_{\text{image}} = \mathcal{E}(I^i_{\text{image}})$, while side information is first converted to textual form and then encoded, yielding $\mathcal{H}^i_{\text{side}} = \mathcal{E}(I^i_{\text{side}})$. These modality-specific features are fused through Q-Former~\citep{li2023blip}, denoted $\mathcal{F}$, to produce the unified item representation:
\begin{equation}
\mathcal{H}^i = \mathcal{F}\left(\mathcal{E}(I^i_{\text{text}}),\; \mathcal{E}(I^i_{\text{image}}),\; \mathcal{E}(I^i_{\text{side}})\right)
\end{equation}

For training, we use item $i^+$ as a positive sample and other items within the same batch as negatives $i^-$, optimizing an InfoNCE-based loss that operates at both the fused and modality-specific levels:
\begin{equation}
\mathcal{L}_{\text{InfoNCE}} = f(\mathcal{H}^i, \mathcal{H}^{i^+}, \mathcal{H}^{i^-}) + f(\mathcal{H}^i_{\text{text}}, \mathcal{H}^{i^+}_{\text{text}}, \mathcal{H}^{i^-}_{\text{text}}) + f(\mathcal{H}^i_{\text{image}}, \mathcal{H}^{i^+}_{\text{image}}, \mathcal{H}^{i^-}_{\text{image}})
\end{equation}
where $f$ denotes the InfoNCE loss function~\citep{radford2021learning}. This contrastive objective is what makes the subsequent quantization stable: the well-separated embedding space it produces prevents the codebook collapse to which RQ-VAE is otherwise prone.

\paragraph{Residual Quantization.}
Given the multimodal item representation $\mathcal{H}^i$, we apply RQ-VAE to map each item to a sequence of discrete codes $\{c_1, c_2, \ldots, c_L\}$, which constitute its Semantic ID. In our final configuration, we employ a two-level codebook structure where each level has a vocabulary size of 32,768. This results in a total of 65,536 new tokens (denoted as \texttt{<C\_0>} through \texttt{<C\_65535>}) being added to the model's vocabulary. A complete SID for an item consists of two codebook entries, e.g., \texttt{<C\_18921><C\_40222>}, where the first token represents the coarse-grained semantic cluster and the second provides fine-grained discrimination within that cluster. This two-level design keeps the added vocabulary size manageable while preserving fine-grained item distinctions: because related items share their coarse code, semantic similarity in the item space surfaces as shared prefixes in SID space, a structure the model can exploit to generalize across related items.

\subsection{Pre-training}
\label{sec:foundation:cpt}
The new SID tokens are appended to the original Qwen3-14B vocabulary, with their embeddings randomly initialized and learned during training. To teach the model to understand and generate this hybrid vocabulary, we adopt a two-stage pipeline: \textbf{Continual Pre-Training} (\S\ref{sec:pretraining_data}), which grounds the SID tokens in item semantics, followed by \textbf{Instruction Tuning} (\S\ref{sec:instruction_tuning}), which teaches the model to apply them across diverse recommendation tasks. A central concern throughout is injecting domain knowledge without eroding the model's pre-existing general capabilities, a balance we strike through careful data construction and strategic mixing.

\subsubsection{Continual Pre-training}
\label{sec:pretraining_data}

Continual pre-training pursues two objectives: {\large\ding{182}}~grounding the newly added SID tokens in natural-language space so the model can understand and generate them in context, and {\large\ding{183}}~retaining the backbone's pre-existing general capabilities. Our corpus therefore combines two complementary components: \textbf{SID-grounding data}, which associates each SID with the content of the item it denotes, teaching the model to read item semantics directly from these tokens, and \textbf{general-domain data}, which guards against catastrophic forgetting.

\paragraph{SID-Grounding Data.}
We construct each SID-grounding sample as a natural-language statement that pairs an item's Semantic ID with its textual attributes, such as title and category, so the model learns to recover an item's content from its SID alone. For example, the following sample associates the SID \texttt{<C\_18921><C\_40222>} with the item's title and category:

\begin{tcolorbox}[
  enhanced,
  colback=observecolor!8,
  colframe=observecolor!75!black,
  boxrule=0.6pt,
  arc=3pt,
  left=10pt, right=10pt, top=8pt, bottom=8pt,
  title={\textbf{Example: SID-Grounding Sample}},
  coltitle=white,
  colbacktitle=observecolor!75!black,
  fonttitle=\small\bfseries
]
\small
The item with Semantic ID \colorbox{hlcolor!12}{\texttt{<C\_18921>}}\,\colorbox{hlcolor!12}{\texttt{<C\_40222>}} has the following information:\\[3pt]
\textbf{Title:} Insulated Lock Rod Insulated Construction Scaffolding Electrified Insulation High Voltage Electric Power Frame.\\[2pt]
\textbf{Category:} Scaffolding.
\end{tcolorbox}
This stage gives the model a working semantics for the SID tokens, so that instruction tuning can operate on identifiers it already understands rather than opaque symbols.

\paragraph{General-Domain Data.}
Training predominantly on SID-grounding data risks catastrophic forgetting of the general-purpose competence that downstream deployment relies on. We therefore blend approximately \textbf{10\% general-domain text} into the corpus, spanning mathematical reasoning, code, scientific literature, medical text, and instruction-following data.

\subsubsection{Instruction Tuning}
\label{sec:instruction_tuning}

Following continual pre-training, we perform instruction tuning to teach the model to apply SID tokens across the diverse alignment tasks. We organize these tasks into three categories: {\large\ding{182}}~bidirectional translation between SIDs and text, {\large\ding{183}}~sequential recommendation carried out directly in SID space, and {\large\ding{184}}~general-domain tasks that preserve the model's pre-existing capabilities. Table~\ref{tab:it_tasks} summarizes the SID-related tasks of the first two categories. In the following, we detail each task type in detail.

\paragraph{SID-Text Bidirectional Translation.}
This family maps SIDs to and from their textual descriptions, linking item titles (\txtp{title}\,$\leftrightarrow$\,\sidp), search queries (\txtp{tag}\,$\leftrightarrow$\,\sidp), and commodities (\sidp\,$\to$\,\txtp{cmd}). Covering both directions teaches the model to read an item's semantics from its SID and, conversely, to emit the right SID for a given description.

\paragraph{Sequential Recommendation.}
The \sidp\,$\to$\,\sidp task casts sequential recommendation fully within the SID space. Given a user's click history organized by multi-granularity temporal windows (\textit{e.g.}, 3-day, 2-day, 1-day, 12-hour, 1-hour), the model predicts the SIDs the user will click next. Operating purely on SID tokens with no textual item descriptions, it provides the most direct evidence that the model has internalized collaborative-filtering signals through these tokens.

\paragraph{General-Domain Tasks.}
To keep the model's general text abilities intact, roughly \textbf{20\% of the instruction tuning data} consists of open-domain supervised tasks such as natural language reasoning, entity extraction, and e-commerce-related tasks. This share is calibrated to preserve instruction-following fidelity, particularly producing well-formed JSON outputs and complying with complex multi-constraint prompts, without diluting the recommendation-specific training signal.

\begin{table}[t]
\centering
\caption{SID-related alignment tasks and their input--output mappings. The upper block covers SID--text bidirectional translation; the bottom row is SID-based sequential recommendation.}
\label{tab:it_tasks}
\renewcommand{\arraystretch}{1.2}
\begin{tabularx}{\textwidth}{@{}l l L c@{}}
\toprule
\textbf{Task} & \textbf{Mapping} & \textbf{Description} & \textbf{Proportion} \\
\midrule
\texttt{sid2title} & \sidp\,$\to$\,\txtp{title} & Generate the item title given an SID. & 13.8\% \\
\texttt{title2sid} & \txtp{title}\,$\to$\,\sidp & Predict the SID given an item title. & 14.7\% \\
\texttt{sid2tag}   & \sidp\,$\to$\,\txtp{tag}   & Generate a natural-language tag given an SID. & 11.5\% \\
\texttt{tag2sid}   & \txtp{tag}\,$\to$\,\sidp   & Predict candidate SIDs given a text tag. & 6.2\% \\
\texttt{sid2cmd}   & \sidp\,$\to$\,\txtp{cmd}   & Predict the commodity category given an SID. & 13.8\% \\
\midrule
\texttt{sid2sid}   & \sidp\,$\to$\,\sidp        & Predict next-click SIDs given a user's click history. & 20.0\% \\
\bottomrule
\end{tabularx}
\end{table}

\section{Latent Intent Reasoning}
\label{sec:reasoning}

Explicit chain-of-thought reasoning has become a common recipe for LLM-based recommenders to interpret user behavior and surface implicit interests, effectively improving prediction accuracy through multi-step deliberation. Yet this quality comes at a steep price: autoregressively generating a rationale of thousands of tokens, far longer than the final output, incurs latency and compute costs that become prohibitive at billion-user scale and high QPS. This raises a natural question: \textbf{\emph{can the gains of deliberative reasoning be retained without paying its token cost?}} 

\begin{figure}[t]
\centering
\includegraphics[width=\textwidth]{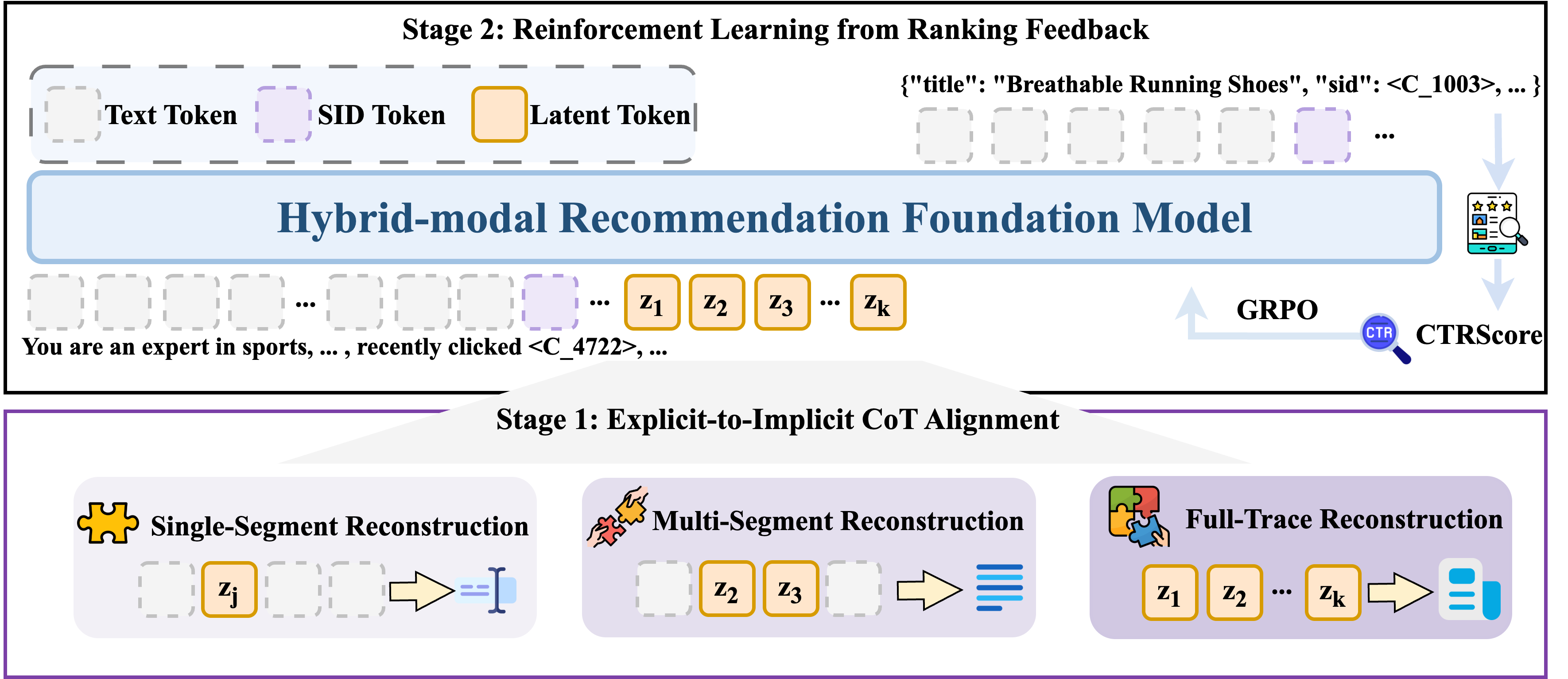}
\caption{Overview of Latent Intent Reasoning. Three reconstruction-based alignment tasks compress an explicit chain-of-thought trace into a short sequence of learnable latent tokens $z$ that faithfully encode it. A two-stage post-training pipeline then internalizes this reasoning into the model: \emph{Explicit-to-Implicit CoT Alignment} first distills teacher traces into the latent tokens, and \emph{Reinforcement Learning from Ranking Feedback} then refines the policy against online business rewards.}
\label{fig:reasoning_overview}
\end{figure}

We answer this question with \textbf{Latent Intent Reasoning}, which internalizes the verbose trace into a short sequence of learnable latent tokens with two properties:
\begin{itemize}
    \item \prop{Efficient}: compressing a rationale of thousands of sequentially decoded steps into a handful of latent tokens shifts the cost from slow autoregressive decoding to parallelizable prefill, reducing reasoning token cost by \hlnum{200$\bf{\times}$} and sharply lowering inference latency.
    \item \prop{Explainable}: unlike opaque latent representations, these tokens can be decoded back into a faithful, human-readable rationale whenever needed, preserving the interpretability that latent reasoning typically sacrifices for speed.
\end{itemize}
To endow these supervision-free tokens with reasoning semantics, we first formulate the latent reasoning mechanism: a compression scheme that condenses a lengthy explicit trace into a few latent tokens, together with multi-task alignment objectives that supervise these tokens to faithfully encode the reasoning they replace (\S\ref{sec:reasoning:efficient}). We then realize this internalization through a complete two-stage post-training pipeline that distills reasoning from a strong teacher and refines it with reinforcement learning driven by online feedback (\S\ref{sec:reasoning:posttraining}). Figure~\ref{fig:reasoning_overview} provides an overview.

\subsection{Reasoning Internalization}
\label{sec:reasoning:efficient}

The reasoning model takes as input $x$, the \emph{persona} assigned by the Global Planner together with the user's recent click history, in which each clicked item is represented by its Semantic ID and short title. Conditioned on $x$, an explicit-CoT model (\textit{e.g.}, DeepSeek-V3.2~\citep{liu2024deepseek} in our experiments) first generates a reasoning trace $R$ of $N$ newline-delimited steps and then the output item tags $y$, factorizing the joint distribution as
\begin{equation}
p_\theta(\textcolor{teal!65!black}{\bm{R}}, y \mid x) = p_\theta(\textcolor{teal!65!black}{\bm{R}} \mid x)\; p_\theta(y \mid x, \textcolor{teal!65!black}{\bm{R}}).
\label{eq:explicit_gen}
\end{equation}
Yet the trace $R$ that carries these reasoning gains is also where the cost concentrates: its $N$ steps are decoded autoregressively and typically far exceed the output $y$ in length, making $R$ the dominant efficiency bottleneck. We therefore internalize $R$ into a short learnable latent slots $z = (z_1, \ldots, z_K)$ with $K \ll N$, introduced as new vocabulary tokens whose embeddings are randomly initialized. This replaces the explicit factorization in Eq.~\eqref{eq:explicit_gen} with its latent counterpart,
\begin{equation}
p_\theta(\textcolor{orange!85!black}{\bm{z}}, y \mid x) = p_\theta(\textcolor{orange!85!black}{\bm{z}} \mid x)\; p_\theta(y \mid x, \textcolor{orange!85!black}{\bm{z}}).
\label{eq:latent_gen}
\end{equation}
It remains to specify how $z$ relates to the implicit derivation trace. We adopt a deterministic positional partition: the trace $R$ is split into $K$ contiguous, non-overlapping segments $R_1, R_2, \ldots, R_K$ of at most $C$ steps each, with latent token $z_j$ encoding segment $R_j$. The number of latent tokens follows:
\begin{equation}
K = \min\!\Bigl(\Bigl\lceil \frac{N}{C} \Bigr\rceil,\; K_{\max}\Bigr)
\label{eq:cot_K}
\end{equation}
Here $C$ sets the granularity of each latent token, with a larger $C$ folding more reasoning steps into a single token, while $K_{\max}$ limits the sequence length and thereby bounds inference cost. When the trace spans fewer than $C \cdot K_{\max}$ steps, the trailing latent tokens cover shorter or empty segments; when it exceeds this budget, coverage stops at the first $K_{\max}$ segments.

\definecolor{cottokbg}{HTML}{DCEAE4}
\definecolor{cotrowA}{HTML}{F3F1F7}
\definecolor{cotrowB}{HTML}{E5E0EF}
\definecolor{cotrowC}{HTML}{D0C7E2}
\newcommand{\cottok}[1]{\colorbox{cottokbg}{\footnotesize\ttfamily<cot#1>}}
\newcommand{\nltext}[1]{\textit{#1}}
\definecolor{xytokbg}{HTML}{D8CFC0}
\newcommand{\xytok}[1]{\colorbox{xytokbg}{\footnotesize$#1$}}
\begin{table}[t]
\centering
\caption{The three alignment tasks on a running example (a badminton-equipment expert), each a choice of the masked set $\mathcal{J}$. Latent tokens \protect\cottok{} replace the segments they encode, and the model reconstructs the masked segments $R_{\mathcal{J}}$ from the surrounding context. The gray anchors \protect\xytok{x} and \protect\xytok{y} mark the input and output, whose specific content is omitted here.}
\label{tab:alignment_tasks}
\footnotesize
\setlength{\fboxsep}{2pt}
\renewcommand{\arraystretch}{1.25}
\begin{tabularx}{\textwidth}{
  >{\raggedright\arraybackslash}p{2.2cm}
  >{\hsize=1.10\hsize\raggedright\arraybackslash}X
  >{\hsize=0.90\hsize\raggedright\arraybackslash}X
}
\toprule
\textbf{Task} & \textbf{Input: context $c(\mathcal{J})$} & \textbf{Target: masked segments $R_{\mathcal{J}}$} \\
\midrule
\rowcolor{cotrowA}
Single-Segment\newline {\small$\mathcal{J}=\{2\}$} &
\xytok{x}\; \nltext{Serious badminton player who favors full-carbon rackets and premium grips.}\; \cottok{2}\; \nltext{Keep the 10 most specific tags and drop broad category terms.}\; \xytok{y} &
\cottok{2}: \nltext{From recent clicks, enumerate rackets, strings, grips, and court shoes.} \\
\rowcolor{cotrowB}
Multi-Segment\newline {\small$\mathcal{J}=\{1,3\}$} &
\xytok{x}\; \cottok{1}\; \nltext{From recent clicks, enumerate rackets, strings, grips, and court shoes.}\; \cottok{3}\; \xytok{y} &
\cottok{1}: \nltext{Serious badminton player who favors full-carbon rackets and premium grips.}\quad \cottok{3}: \nltext{Keep the 10 most specific tags and drop broad category terms.} \\
\rowcolor{cotrowC}
Full-Trace\newline {\small$\mathcal{J}=\{1,2,3\}$} &
\xytok{x}\; \cottok{1}\,\cottok{2}\,\cottok{3}\; \xytok{y} &
\nltext{Serious badminton player who favors full-carbon rackets and premium grips. From recent clicks, enumerate rackets, strings, grips, and court shoes. Keep the 10 most specific tags and drop broad category terms.} \\
\arrayrulecolor{black}
\bottomrule
\end{tabularx}
\end{table}

\paragraph{Latent Token Warm-up.}
The randomly initialized latent embeddings lag far behind the pretrained token embeddings: they lie outside the pretrained embedding distribution and carry no semantics yet. To calibrate them into a representation space compatible with the language tokens, we randomly select one or more contiguous spans of the trace $R$, replace each with latent tokens to form a mixed trace $\tilde{R}$, and optimize the standard autoregressive objective over the mixed sequence $w = (x, \tilde{R}, y)$,
\begin{equation}
\mathcal{L}_{\text{warm}} = -\sum_{t \in \mathcal{T}} \log p_\theta\!\left(w_t \mid w_{<t}\right),
\label{eq:warmup_lm}
\end{equation}
where $\mathcal{T}$ indexes the surviving text tokens and the latent positions contribute only as context. Predicting each masked span from the surrounding text pulls the latent embeddings into the pretrained representation space and endows them with coarse contextual semantics, yielding a well-conditioned initialization for the subsequent segment-level alignment.

\paragraph{Multi-granularity Alignment.}
The warm-up stage calibrates the latent embeddings at the representation level, but it does not determine what information each latent slot should preserve. The positional partition assigns segment $R_j$ to latent token $z_j$ only as a structural correspondence; this correspondence must be converted into a learnable information constraint. We instantiate this constraint through masked reconstruction at multiple granularities. Given a masked index set $\mathcal{J} \subseteq \{1, \ldots, K\}$, we form a mixed sequence that keeps unmasked segments as text while replacing each masked one with its latent token, which is formulated as follows:
\begin{equation}
c(\mathcal{J}) = \bigl(x,\; e_1, \ldots, e_K,\; y\bigr), \qquad
e_j =
\begin{cases}
z_j, & j \in \mathcal{J},\\
R_j, & j \notin \mathcal{J},
\end{cases}
\label{eq:mixed_context}
\end{equation}
and, cued by a task-specific instruction prompt $\mathcal{P}$ that indicates which positions to recover, train the model to reconstruct the masked segments $R_{\mathcal{J}} = \{R_j\}_{j \in \mathcal{J}}$ from this context:
\begin{equation}
\mathcal{L}(\mathcal{J}) = -\log p_\theta\bigl(R_{\mathcal{J}} \mid c(\mathcal{J}),\, \mathcal{P}\bigr).
\label{eq:recon_loss}
\end{equation}
The prompt template for each task is provided in Appendix~\ref{app:reasoning_prompts}.
The three tasks differ only in how $\mathcal{J}$ is drawn, widening from a single latent token to the full sequence (as illustrated in Table~\ref{tab:alignment_tasks}):
\begin{itemize}
\item \textbf{Single-Segment Reconstruction} ($\mathcal{J} = \{j\}$). We hide a single segment $R_j$ behind its latent token $z_j$ while leaving the rest of the trace in text, and the model reconstructs $R_j$. The rich surrounding context makes this reconstruction straightforward, so minimizing $\mathcal{L}(\{j\})$ grounds each latent token in its own segment and yields a direct per-position encoding signal.
\item \textbf{Multi-Segment Reconstruction} ($\mathcal{J} \subsetneq \{1, \ldots, K\}$, $|\mathcal{J}| \geq 2$). We hide several segments behind their latent tokens simultaneously, and the model restores only the masked text $R_{\mathcal{J}}$. As the masked spans may straddle segment boundaries, minimizing $\mathcal{L}(\mathcal{J})$ drives each token to remain informative alongside the surrounding text rather than in isolation.
\item \textbf{Full-Trace Reconstruction} ($\mathcal{J} = \{1, \ldots, K\}$). We mask every segment, so the context reduces to $c = (x, z, y)$ and the objective becomes $\mathcal{L}(\{1, \ldots, K\}) = -\log p_\theta(R \mid x, z, y, \mathcal{P})$. This compels the latent sequence to be sufficient to recover the full reasoning chain from $(x, z, y)$ alone.
\end{itemize}

Beyond compression, the reconstruction objective preserves the decodability of latent reasoning. Because $z$ is trained as a sufficient encoding of $R$, the model can recover a human-readable rationale from $(x, z, y)$, retaining the transparency of explicit CoT at a much lower token cost. Appendix~\ref{app:reasoning_prompts} provides cases showing how the latent tokens correspond to the explicit reasoning chain.

\subsection{Post-training}
\label{sec:reasoning:posttraining}

This section details the post-training pipeline that equips \recgpt[-V3] with latent reasoning, carried out in two stages: (1)~\textbf{Explicit-to-Implicit CoT Alignment} (\S\ref{sec:reasoning:posttraining:alignment}) distills reasoning from a strong teacher into explicit traces and then compresses them into latent tokens through the multi-task curriculum of \S\ref{sec:reasoning:efficient}; (2)~\textbf{Reinforcement Learning from Ranking Feedback} (\S\ref{sec:reasoning:posttraining:rl}) optimizes the latent model directly against online business objectives.

\subsubsection{Stage 1: Explicit-to-Implicit CoT Alignment}
\label{sec:reasoning:posttraining:alignment}

In this training stage, we first distill reasoning from a strong teacher into explicit CoT traces and then compress them into latent tokens through the multi-task curriculum of \S\ref{sec:reasoning:efficient}.

\begin{wraptable}[7]{r}{0.43\textwidth}
\centering
\vspace{-1.5em}
\caption{Training data mixture in Stage~1.}
\vspace{-0.7em}
\label{tab:cot_training}
\footnotesize
\renewcommand{\arraystretch}{1.1}
\begin{tabular}{lr}
\toprule
\textbf{Task} & \textbf{Prop.} \\
\midrule
\rowcolor{teal!18}
\ding{182}~\textbf{Reasoning alignment} & 21.43\% \\
\rowcolor{teal!18}
\quad Single-segment recon. & \phantom{0}6.45\% \\
\rowcolor{teal!18}
\quad Multi-segment recon. & \phantom{0}6.56\% \\
\rowcolor{teal!18}
\quad Full-trace recon. & \phantom{0}8.42\% \\
\rowcolor{gray!16}
\ding{183}~\textbf{General reasoning} & 69.58\% \\
\rowcolor{orange!18}
\ding{184}~\textbf{Tag prediction} & \phantom{0}8.43\% \\
\bottomrule
\end{tabular}
\end{wraptable}

\paragraph{(1) Reasoning Distillation.}
We distill reasoning from a strong teacher by fine-tuning the model on its explicit CoT outputs. These thinking processes are effective yet inefficient, averaging roughly 2{,}300 tokens and thus dominating overall inference cost. 

\paragraph{(2) Latent Internalization.}
We then compress each trace into latent tokens via the alignment curriculum of \S\ref{sec:reasoning:efficient}; with $C=20$ and $K_{\max}=10$, a trace maps to at most 10 latent tokens, a roughly 200:1 reduction. Training draws on a multi-task mixture (Table~\ref{tab:cot_training}) spanning three roles: {\large\ding{182}}~the three \textbf{Reasoning-Alignment} tasks that internalize the trace into latent tokens, {\large\ding{183}}~\textbf{General Reasoning} data that guards against catastrophic forgetting, and {\large\ding{184}}~\textbf{Tag Prediction} that matches the deployment format, on which we supervise only the output tokens to spare the latent positions from collapse.

\subsubsection{Stage 2: Reinforcement Learning from Ranking Feedback}
\label{sec:reasoning:posttraining:rl}

The supervised stages only imitate teacher trajectories and cannot optimize business objectives. Existing RecGPT-V2~\citep{yi2025recgptv2} adopts an accuracy reward based on \metric{HitRate}, the fraction of a user's held-out ground-truth items recovered among the candidates its outputs retrieve, yet this offline proxy exhibits two limitations. \textbf{(1) \emph{Reward sparsity}}: under a GRPO-style objective, rollouts within a group can receive the same reward, collapsing their advantages to zero and yielding no learning gradient. \textbf{(2) \emph{Pipeline inconsistency}}: since outputs must traverse the serving pipeline before reaching users, a reward computed outside it may diverge from actual downstream performance.

To address both limitations, we propose \textbf{Reinforcement Learning from Ranking Feedback (RLRF)}, whose core idea is to draw the reward directly from the production ranking model that governs what users ultimately see, yielding a signal that is both dense and consistent with the serving pipeline. Specifically, we replace the \metric{HitRate} accuracy reward with a \metric{CTRScore} read from this downstream ranking model. For a generated output $y = \{t_1, \ldots, t_m\}$ of $m$ item tags, we retrieve candidate items with $y$, score each with the production ranking model, and average the top-$K$ scores:
\begin{equation}
r_{\text{ctr}}(y) = \frac{1}{K} \sum_{k=1}^{K} s_k,
\label{eq:ctr_score}
\end{equation}
where $s_k$ is the $k$-th largest \metric{CTRScore} among the retrieved items ($K=100$ in our experiments). This design counters both limitations: averaging over the top-$K$ ranking scores turns the sparse \metric{HitRate} signal into a dense one, while reading the reward from the production ranking model aligns training with the pipeline the outputs actually traverse. An output thus earns a high reward precisely when the items it retrieves are favored by the downstream ranker, steering the policy toward candidates the pipeline is more likely to surface to users.

For the remaining reward terms, we retain the \emph{\textbf{Constrained Reward Shaping (CRS)}} framework of RecGPT-V2, which treats an \emph{alignment score}, a \emph{diversity score}, and a \emph{length score} as quality constraints on the primary reward: $r_{\text{ctr}}$ propagates only when all three thresholds are met. 
The alignment score is produced by a reward model trained on human-preference pairs, rating how well each predicted tag meets human quality standards. 
The final reward is formulated as follows:
\begin{equation}
\mathcal{R}(y) = r_{\text{ctr}}(y)\,\mathbbm{1}\!\left[\text{align}(y) \geq \tau_{\text{align}}\right]\mathbbm{1}\!\left[\text{div}(y) \geq \tau_{\text{div}}\right]\mathbbm{1}\!\left[\text{len}(y) \geq \tau_{\text{len}}\right],
\label{eq:crs_reward}
\end{equation}
where $\tau_{\text{align}}$, $\tau_{\text{div}}$, and $\tau_{\text{len}}$ are the alignment, diversity, and length thresholds; we refer readers to RecGPT-V2 for the definitions of these terms. We optimize the policy with GRPO~\citep{shao2024deepseekmath}, computing group-relative advantages over a group of outputs sampled per input.

\section{Experiments}
\label{sec:experiments} 
We evaluate RecGPT-V3 on Taobao's homepage recommendation scenario, serving hundreds of millions of daily active users. Our evaluation covers five dimensions: (1)~overall online A/B test results against the RecGPT-V2 baseline (\S\ref{sec:exp:abtest}); (2)~memory hub evaluation through human annotation and compute efficiency analysis (\S\ref{sec:exp:memory}); (3)~recommendation foundation model evaluation on general-capability preservation, SID-text alignment, and downstream recommendation quality (\S\ref{sec:exp:foundation}); (4)~latent reasoning evaluation through post-training effectiveness and inference efficiency analysis (\S\ref{sec:exp:reasoning}); and (5)~further analysis on SID modality and case studies (\S\ref{sec:exp:further}).

\subsection{Online A/B Test Results}
\label{sec:exp:abtest}
\paragraph{Experimental Setup.}
We deploy RecGPT-V3 in Taobao's homepage ``Guess What You Like'' scenario and conduct online A/B tests under the following configuration:
\begin{itemize}
    \item \textbf{Traffic Allocation:} The experimental and control groups each receive 1\% of total platform traffic, yielding statistically significant results while keeping deployment risk to a minimum.
    \item \textbf{Baseline:} RecGPT-V2 serves as the control group, enabling a direct assessment of the improvements introduced by RecGPT-V3.
    \item \textbf{Evaluation Scenarios:} We evaluate performance separately under two complementary scopes:
    \begin{itemize}
        \item \emph{Item Scenario:} Direct item recommendations presented in a grid layout.
        \item \emph{Feed Scenario:} A mixed-content recommendation stream in the main feed, comprising items, advertisements, live streams, and other content types.
    \end{itemize}
\end{itemize}
 
\paragraph{Evaluation Metrics.}
We assess system performance along two dimensions:

\textbf{User Engagement:}
\begin{itemize}[topsep=2pt, itemsep=2pt]
    \item \prop{IPV (Item Page Views):} Number of item detail page visits, reflecting user interest.
    \item \prop{CTR (Click-Through Rate):} Ratio of clicks to impressions, measuring recommendation relevance.
    \item \prop{PV (Page Views):} Total number of page views, indicating overall browsing activity.
    \item \prop{DAU (Daily Active Users):} Number of unique active users per day, measuring platform activity.
\end{itemize}

\textbf{Business Transaction:}
\begin{itemize}[topsep=2pt, itemsep=2pt]
    \item \prop{TC (Transaction Count):} Number of completed transactions, reflecting conversion effectiveness.
    \item \prop{GMV (Gross Merchandise Value):} Total monetary value of transactions, measuring overall business value.
\end{itemize}
 
\begin{table}[t]
\centering
\caption{Online A/B test results comparing RecGPT-V3 against RecGPT-V2 baseline across item and feed scenarios. All metrics show relative percentage improvements (\% omitted).}
\label{tab:online-ab}
\begin{tabular}{l cccc cc}
\toprule
\multirow{2}{*}{\textbf{Scenario}} & \multicolumn{4}{c}{\textbf{User Engagement}} & \multicolumn{2}{c}{\textbf{Business Transaction}} \\
\cmidrule(lr){2-5} \cmidrule(lr){6-7}
 & IPV & CTR & PV & DAU & TC & GMV \\
\midrule 
Item & +3.08 & +0.98 & +2.02 & -- & +3.10 & +7.51 \\
Feed & +1.28 & +1.00 & +0.83 & +0.56 & +1.97 & +3.97 \\
\bottomrule
\end{tabular}

\vspace{2pt}
{\small Note: -- indicates metrics not applicable in the given scenario.}
\end{table}
 
\paragraph{Results and Analysis.}
Table~\ref{tab:online-ab} summarizes the online A/B test results. RecGPT-V3 consistently outperforms RecGPT-V2 across all metrics in both evaluation scenarios.

$\blacklozenge$\; In the item scenario, the improvements are more pronounced: +3.08\% IPV, +0.98\% CTR, +2.02\% PV, +3.10\% TC, and +7.51\% GMV. This pattern suggests that RecGPT-V3 improves item-level recommendation beyond simply attracting more clicks, as the relatively stronger gains in IPV, TC, and especially GMV indicate that users are more likely to enter product detail pages, complete transactions, and generate higher transaction value after exposure. This implies the retrieved candidates are not only more relevant at the click stage, but also better aligned with users' purchase-oriented intent.
 
$\blacklozenge$\; In the feed scenario, which reflects the overall platform experience, RecGPT-V3 delivers +1.28\% IPV and +1.00\% CTR, indicating improved recommendation relevance and user engagement. Browsing activity rises by +0.83\% PV, and platform retention shows a healthy gain of +0.56\% DAU, confirming that the improvements extend across a broad user base rather than concentrating on a narrow subset. Transaction metrics also improve substantially, with GMV increasing by +3.97\% and TC by +1.97\%, demonstrating that stronger intent understanding translates directly into purchase activity.

\subsection{Memory Hub Evaluation}
\label{sec:exp:memory}
We evaluate the memory hub introduced in \S\ref{sec:memory} through two complementary analyses: human annotation to validate memory representation accuracy (\S\ref{sec:exp:memory:human}) and compute efficiency analysis (\S\ref{sec:exp:memory:compute}). In all experiments, the memory hub condition replaces the Global Planner's full-sequence input with the compressed memory representation plus the recent behavioral delta, while maintaining identical downstream components.

\subsubsection{Memory Quality Assessment}
\label{sec:exp:memory:human}

\begin{table}[t]
\centering
\caption{Human evaluation of memory unit quality. ``Behavior Pattern'' measures whether the assigned pattern identifier correctly categorizes the user's behavioral cluster; ``Behavior Index'' measures whether the representative indices correctly point to interactions belonging to the identified pattern.}
\label{tab:memory-human}
\begin{tabular}{lrc}
\toprule
\textbf{Evaluation Target} & \textbf{Annotations} & \textbf{Accuracy} \\
\midrule
Behavior Pattern & 2,514  & 82.89\% \\
Behavior Index   & 21,268 & 95.27\% \\
\bottomrule
\end{tabular}
\end{table}

We conduct a large-scale human annotation study on the memory units produced by Structured Behavior Compression (\S\ref{sec:sbc}) to evaluate the factual accuracy of the memory hub's structured representations. Annotators assess two complementary aspects of each memory unit, as shown in Table~\ref{tab:memory-human}.
Across 2,514 annotated behavior patterns, accuracy reaches 82.89\%, indicating that the LLM-based compression reliably identifies coherent behavioral archetypes from noisy interaction sequences. Behavior index accuracy is substantially higher at 95.27\% across 21,268 annotations, confirming that once a pattern is correctly identified, the supporting evidence is accurately attributed. This is particularly important because each memory unit stores only integer pointers into the original sequence (\S\ref{sec:sbc}), and the high index accuracy ensures these pointers remain faithful references for downstream reasoning. Together, the two metrics validate that the memory hub produces reliable structured representations.

\subsubsection{Compute Efficiency}
\label{sec:exp:memory:compute}

\begin{wraptable}{r}{0.5\textwidth}
\centering
\vspace{-15pt}
\small
\caption{User-modeling compute cost with and without the memory hub, expressed relative to the RecGPT-V2 baseline.}
\label{tab:compute}
\begin{tabular}{@{}llc@{}}
\toprule
\textbf{System} & \textbf{Component} & \textbf{Cost} \\
\midrule
RecGPT-V2 & Full sequence & $100\%$ \\
\midrule
\multirow{3}{*}{RecGPT-V3} & Per-inference & $33.43\%$ \\
                           & Memory curation & $10.77\%$ \\
\cmidrule(l){2-3}
                           & \textbf{Total} & \textbf{44.20\%} \\
\bottomrule
\end{tabular}
\vspace{-10pt}
\end{wraptable}

The memory hub reduces the Global Planner's compute cost by \textbf{55.80\%} relative to the RecGPT-V2 baseline (Table~\ref{tab:compute}). The dominant saving comes at inference: each pass now conditions on the compressed memory and the recent behavioral delta rather than re-encoding the full behavioral sequence, bringing the per-pass cost down to \textbf{33.43\%}. Periodic incremental curation of the memory (\S\ref{sec:emc}) contributes a further \textbf{10.77\%}, a modest overhead relative to the saving it enables. Overall, the memory hub converts a recurring per-inference expense into an amortized memory maintenance cost, yielding a compute profile that scales favorably with request volume.

\subsection{Foundation Model Evaluation}
\label{sec:exp:foundation}

We evaluate the hybrid-modal foundation model along two questions: (1) Does the model preserve strong general language capabilities despite domain-specific training (\S\ref{sec:exp:foundation:general})? (2) Does the model achieve accurate cross-modal SID--text alignment and translate it into strong downstream recommendation quality (\S\ref{sec:exp:foundation:alignment})? To disentangle the contribution of general-domain data mixing, we compare two model variants throughout:

\begin{itemize}
    \item \emph{\underline{w/ General-Domain Data}}: Initialized from Qwen3-14B and trained with both SID-grounding data and general-domain data.
    \item \emph{\underline{w/o General-Domain Data}}: Same architecture and SID-grounding data, but with all general-domain data removed.
\end{itemize}

\subsubsection{General Capability Preservation}
\label{sec:exp:foundation:general}

To answer whether the hybrid-modal foundation model preserves general language capabilities, we evaluate its accuracy across four representative benchmarks spanning mathematical reasoning (GSM8K), broad knowledge (MMLU), Chinese language understanding (CMMLU), and instruction following (IFEval).

\begin{figure}[htbp]
\centering
\includegraphics[width=0.75\textwidth]{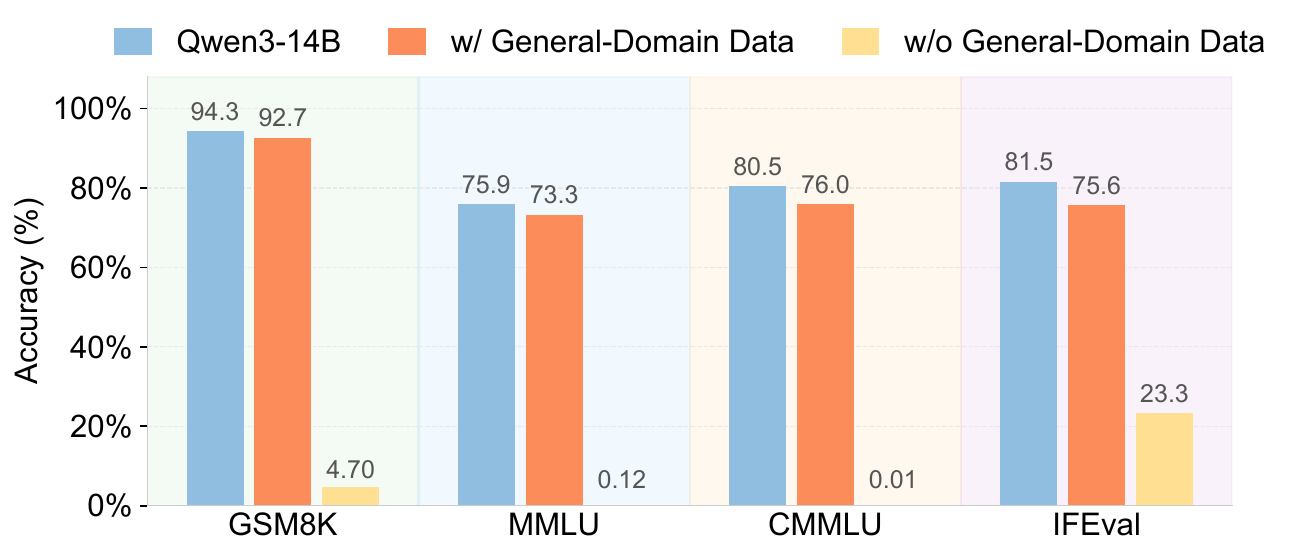}
\caption{General capability evaluation across four benchmarks. The hybrid-modal foundation model (\emph{\underline{w/ General-Domain Data}}) preserves most of the backbone's capabilities, while removing general-domain data (\emph{\underline{w/o General-Domain Data}}) leads to catastrophic collapse.}
\label{fig:general-capability}
\end{figure}

As shown in Figure~\ref{fig:general-capability}, general-domain data mixing preserves the vast majority of the backbone's capabilities. GSM8K drops only 1.66\% (94.31\% $\to$ 92.65\%), MMLU 2.65\%, CMMLU 4.49\%, and IFEval falls from 81.52\% to 75.60\%. Removing general-domain data instead causes catastrophic collapse. The \emph{\underline{w/o General-Domain Data}} variant falls to 4.70\% on GSM8K, 0.12\% on MMLU, 0.01\% on CMMLU, and 23.29\% on IFEval, and loses nearly all reasoning, knowledge, and instruction-following ability. General-domain data is therefore essential. Without the regularization it provides during continual pre-training and instruction tuning, the model overfits to domain-specific patterns and loses the general capabilities needed for real-world deployment.

\subsubsection{SID Alignment and Recommendation Quality}
\label{sec:exp:foundation:alignment}
\label{sec:exp:foundation:downstream}

\begin{wrapfigure}[15]{r}{0.46\textwidth}
\centering
\vspace{-20pt}
\includegraphics[width=0.48\textwidth]{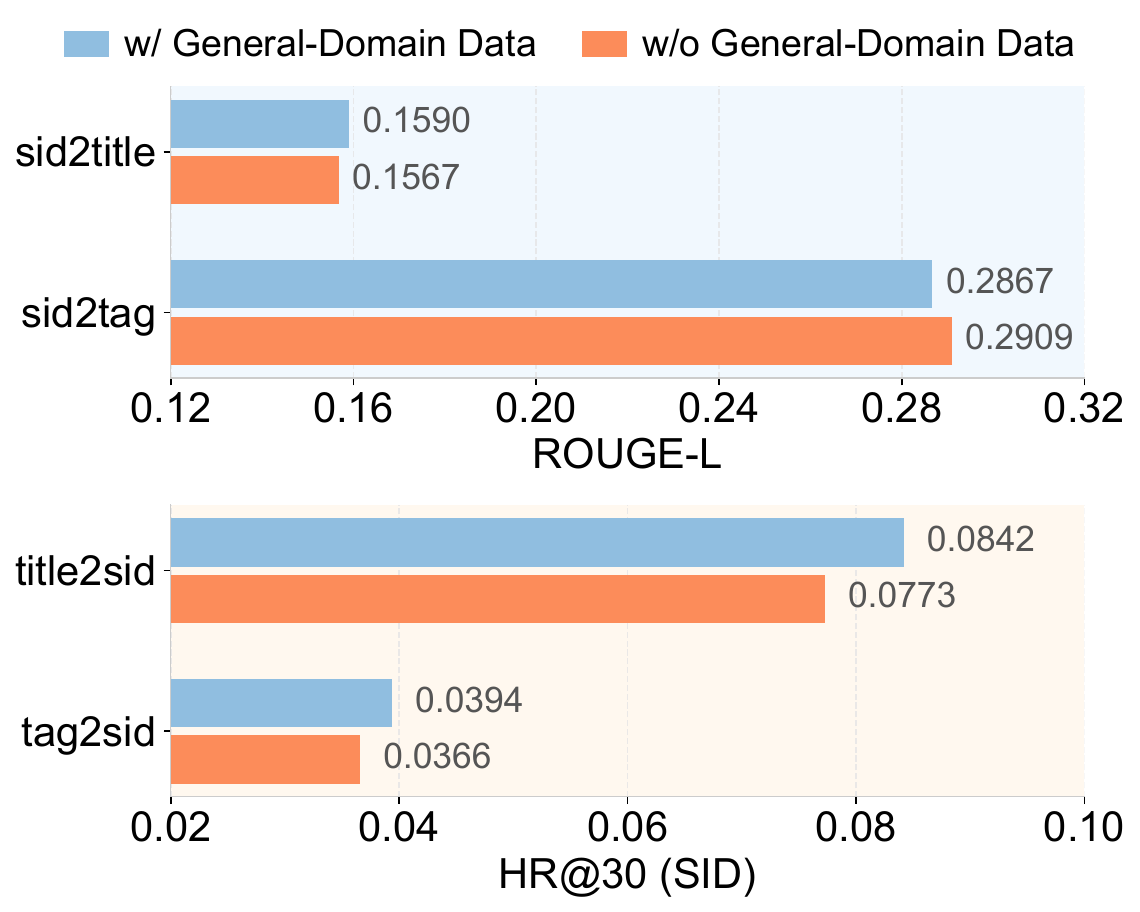}
\vspace{-22pt}
\caption{SID--text semantic alignment across four bidirectional translation tasks.}
\label{fig:sid-alignment}
\end{wrapfigure}
We assess the foundation model along two coupled dimensions: cross-modal alignment between SIDs and text, and downstream recommendation quality. Alignment is measured across four bidirectional tasks (\emph{sid2title}, \emph{sid2tag}, \emph{title2sid}, \emph{tag2sid}), with ROUGE-L for the two generation tasks and HR@30 (SID) for the two prediction tasks. Downstream quality is measured on three axes: tag diversity (average unique tags per user), category coverage (distinct item categories after mapping tags to the Taobao taxonomy), and category-level HR@30 (hit rate against actual clicks over two weeks).

Beyond mitigating catastrophic forgetting of the backbone's general capabilities (\S\ref{sec:exp:foundation:general}), general-domain data preserves SID--language alignment and lifts downstream recommendation quality. Figure~\ref{fig:sid-alignment} shows that all four alignment metrics remain essentially unchanged between the two variants, indicating that mixing general-domain data does not compromise the SID-to-text mapping ability learned from SID-grounding data. Table~\ref{tab:tag-quality} further shows that this data mixing lifts category-level HR@30 by $26.2\%$ ($0.2250 \to 0.3050$) alongside comparable gains in tag diversity and category coverage, translating the retained general capabilities into materially stronger downstream performance through content-based and context-aware user intent understanding. Overall, these results justify the deliberate integration of general-domain data: it preserves the semantic alignment advantage, retains the backbone's general knowledge, and converts both into stronger downstream performance.

\begin{table}[]
\centering
\caption{Downstream recommendation quality comparison.}
\label{tab:tag-quality}
\begin{tabular}{lccc}
\toprule
{\textbf{Model}} & \textbf{Avg. \#Tags} & \textbf{Avg. \#Categories} & \textbf{HR@30 (Category)} \\
\midrule
\emph{w/ General-Domain Data} & 28.77 & 41.73 & 0.3050 \\
\emph{w/o General-Domain Data} & 23.88 & 32.49 & 0.2250 \\
\bottomrule
\end{tabular}
\end{table}

\subsection{Latent Reasoning Evaluation}
\label{sec:exp:reasoning}

We evaluate the latent reasoning module (\S\ref{sec:reasoning}) from two perspectives: (1)~post-training effectiveness, examining how each training stage and RL contribute to recommendation quality; and (2)~inference efficiency, comparing the computational cost of explicit versus latent reasoning.

\subsubsection{Post-training Effectiveness}
\label{sec:exp:reasoning:ablation}

We progressively add training stages to examine their individual contributions to recommendation quality. Table~\ref{tab:reasoning-results} reports two metrics: HR@30 (Category), the category-level hit rate among top-30 predictions, and CTR, the average CTR of top-100 items retrieved by the model's outputs. We compare two groups of model variants: the first is based on Qwen3-14B, assessing whether native reasoning alone improves recommendation without domain-specific training (CTR is not applicable as these variants are not connected to the online feedback pipeline); the second builds upon the hybrid-modal foundation model (\S\ref{sec:foundation}), progressively adding reasoning components.
\begin{itemize}
    \item \emph{\underline{Qwen3-14B}}: the base language model without reasoning, directly generating outputs from prompts.
    \begin{itemize}
        \item \emph{\underline{+ Native Reasoning}}: enabling Qwen3's built-in chain-of-thought during inference.
    \end{itemize}
    \item \emph{\underline{Hybrid-modal Foundation Model}}: the foundation model after pre-training, without reasoning.
    \begin{itemize}
        \item \emph{\underline{+ Explicit CoT (SFT)}}: supervised fine-tuning with reasoning traces distilled from DeepSeek-V3.2.
        \item \emph{\underline{+ Latent Reasoning}}: internalizing explicit reasoning into 10 latent tokens via the multi-task curriculum (\S\ref{sec:reasoning:efficient}).
        \item \emph{\underline{+ RL}}: applying RL on top of latent reasoning (the full RecGPT-V3).
    \end{itemize}
\end{itemize}

On Qwen3-14B, enabling native chain-of-thought yields only a marginal HR@30 improvement (0.2347 vs.\ 0.2276), confirming that reasoning without domain-specific training provides limited benefit. On the hybrid-modal foundation model, explicit CoT improves HR@30 from 0.3050 to 0.3508, demonstrating that structured reasoning enables more precise intent understanding. Latent reasoning achieves comparable quality (HR@30 0.3462, CTR 0.0649) while compressing $\sim$2{,}700 reasoning tokens into just 10 latent tokens. With RL, performance further improves to 0.3693 on HR@30 and 0.0679 on CTR, surpassing explicit CoT on both metrics.

\begin{table}[t]
\centering
\caption{Post-training effectiveness comparison. The upper group evaluates reasoning on the base language model; the lower group evaluates on the hybrid-modal foundation model. ``--'' indicates metrics not applicable to configurations outside the online feedback pipeline.}
\label{tab:reasoning-results}
\small
\begin{tabular}{@{}lcc@{}}
\toprule
\textbf{Setting} & \textbf{HR@30 (Category)} & \textbf{CTR} \\
\midrule
Qwen3-14B & 0.2276 & -- \\
\quad \emph{+ Native Reasoning} & 0.2347 & -- \\
\midrule
Hybrid-modal Foundation Model & 0.3050 & 0.0624 \\
\quad \emph{+ Explicit CoT (SFT)} & 0.3508 & 0.0638 \\
\quad \emph{+ Latent Reasoning} & 0.3462 & 0.0649 \\
\quad \emph{+ RL} & \textbf{0.3693} & \textbf{0.0679} \\
\bottomrule
\end{tabular}
\end{table}

\subsubsection{Inference Efficiency}

A key practical concern with explicit chain-of-thought is its computational overhead: generating $\sim$2{,}840 reasoning tokens per sample significantly increases inference latency, as each token requires sequential autoregressive decoding. Latent reasoning addresses this by compressing the reasoning process into 10 latent tokens, shifting the computational cost from slow sequential decoding to parallelizable prefill computation.
Table~\ref{tab:inference-efficiency} quantifies this advantage on a 1{,}000-sample benchmark under identical hardware. The latent reasoning model reduces per-sample output length from 2{,}840 to 122 tokens (a 95.7\% reduction), yielding a 3.46$\times$ end-to-end speedup (1{,}020s $\to$ 295s). Meanwhile, input throughput increases from 166K to 498K tokens per minute, confirming that the bottleneck has effectively shifted from output decoding to input prefill. Combined with the comparable recommendation quality demonstrated in Table~\ref{tab:reasoning-results}, latent reasoning achieves a favorable trade-off between effectiveness and efficiency for production deployment.

\paragraph{Serving Cost Analysis.}
Using semantic IDs (SIDs) in both input and output, together with latent intent reasoning, increases the context length and thus introduces a 15\% computational overhead for expert model compared with the RecGPT-V2 expert. However, this local overhead is outweighed by the system-level savings from the Global Planner, whose computational cost is approximately 20 times that of an expert model in the current deployment. With the memory mechanism reducing the planner-side computation by 55.8\%, the weighted overall resource saving reaches $\hlnum{52.4\%}$.

\begin{table}[t]
\centering
\caption{Inference efficiency comparison between explicit CoT and latent reasoning on 1{,}000 samples under identical hardware. ``Output Length'' denotes the average number of tokens generated per sample, including reasoning and final output; ``Input/Output TPM'' denotes the tokens-per-minute throughput during prefill and decoding, respectively; ``Total Time'' denotes the wall-clock time to process all samples.}
\label{tab:inference-efficiency}
\small
\begin{tabular}{@{}lrrrr@{}}
\toprule
\textbf{Mode} & \textbf{Output Length.} & \textbf{Input TPM} & \textbf{Output TPM} & \textbf{Total Time (s)} \\
\midrule
Explicit CoT & 2{,}840 & 166K & 531K & 1{,}020 \\
Latent Reasoning & 122 & 498K & 66.7K & 295 ($\downarrow$71.1\%) \\
\bottomrule
\end{tabular}
\end{table}

\subsection{Further Analysis}
\label{sec:exp:further}

Beyond the quantitative evaluations above, we provide two further analyses to offer deeper insights. We first examine the complementary roles of text tags and SIDs as dual retrieval modalities (\S\ref{sec:exp:sid}), then present case studies illustrating how the core modules collaborate in practice (\S\ref{sec:exp:case}).

\subsubsection{SID Modality Analysis}
\label{sec:exp:sid}

The hybrid-modal foundation model generates both text tags and SIDs as complementary intent representations (\S\ref{sec:foundation}): text tags provide generalizable category-level coverage through open-ended world knowledge, while SIDs ground user interests in concrete collaborative semantics. We validate this design from three perspectives: the coverage breadth of text tags versus the user specificity of SIDs, their diversity in the embedding space, and their complementarity in end-to-end retrieval.

\paragraph{Coverage Breadth vs. User Specificity.}
In each inference, the expert model generates $N$ text tags $\{t_1, t_2, \ldots, t_N\}$ for a given user, along with $K$ SIDs $\{\text{SID}_1, \ldots, \text{SID}_K\}$ per tag. We map each text tag $t_i$ to an item category set $\mathcal{C}_{\text{tag}}(t_i)$, and aggregate its co-generated SID categories as $\mathcal{C}_{\text{sid}}(t_i) = \bigcup_{k=1}^{K} \mathcal{C}(\text{SID}_k)$. We define the average category breadth $\tilde{\mathcal{C}}$ and cross-user overlap $J$ to characterize each signal type:
\begin{equation}
    \tilde{\mathcal{C}}_{\text{tag}} = \frac{1}{N}\sum_{i=1}^{N} |\mathcal{C}_{\text{tag}}(t_i)|, \qquad
    \tilde{\mathcal{C}}_{\text{sid}} = \frac{1}{N}\sum_{i=1}^{N} |\mathcal{C}_{\text{sid}}(t_i)|,
    \label{eq:category-breadth}
\end{equation}
\begin{equation}
    J_{\text{tag}} = \mathbb{E}_{(u,v)}\left[\frac{|\mathcal{C}_{\text{tag}}^{(u)} \cap \mathcal{C}_{\text{tag}}^{(v)}|}{|\mathcal{C}_{\text{tag}}^{(u)} \cup \mathcal{C}_{\text{tag}}^{(v)}|}\right], \qquad
    J_{\text{sid}} = \mathbb{E}_{(u,v)}\left[\frac{|\mathcal{C}_{\text{sid}}^{(u)} \cap \mathcal{C}_{\text{sid}}^{(v)}|}{|\mathcal{C}_{\text{sid}}^{(u)} \cup \mathcal{C}_{\text{sid}}^{(v)}|}\right],
    \label{eq:cross-user-sim}
\end{equation}
\begin{wraptable}{r}{0.42\textwidth}
\centering
\vspace{-10pt}
\caption{Comparison between text tags and SIDs on category-level statistics.}
\label{tab:tag-sid-comparison}
\begin{tabular}{lcc}
\toprule
\textbf{Metric} & \textbf{Text Tag} & \textbf{SID} \\
\midrule
Category Breadth & 1.40 & 0.84 \\
Cross-User Overlap & 11.36\% & 4.61\% \\
\bottomrule
\end{tabular}
\vspace{-10pt}
\end{wraptable}
where $\tilde{\mathcal{C}}$ measures the average number of distinct categories covered, and $J$ measures the expected overlap in category sets between randomly sampled user pairs $(u, v)$. Results are reported in Table~\ref{tab:tag-sid-comparison}.
On category breadth, $\tilde{\mathcal{C}}_{\text{tag}} = 1.40$ while $\tilde{\mathcal{C}}_{\text{sid}} = 0.84$. Even when aggregating $K$ SIDs per tag, the combined category set remains narrower than that of a single text tag. This confirms the broader coverage of text tags, which span diverse interest scopes, against the narrower focus of SIDs on specific category clusters.
On cross-user overlap, $J_{\text{tag}} = 11.36\%$ compared to $J_{\text{sid}} = 4.61\%$. The substantially lower overlap of SIDs reflects their stronger user specificity, as collaborative semantics capture user-specific signals, whereas the higher overlap of text tags stems from their reliance on shared world knowledge.

\begin{wrapfigure}[13]{r}{0.45\textwidth}
\centering
\vspace{-10pt}
\includegraphics[width=0.42\textwidth]{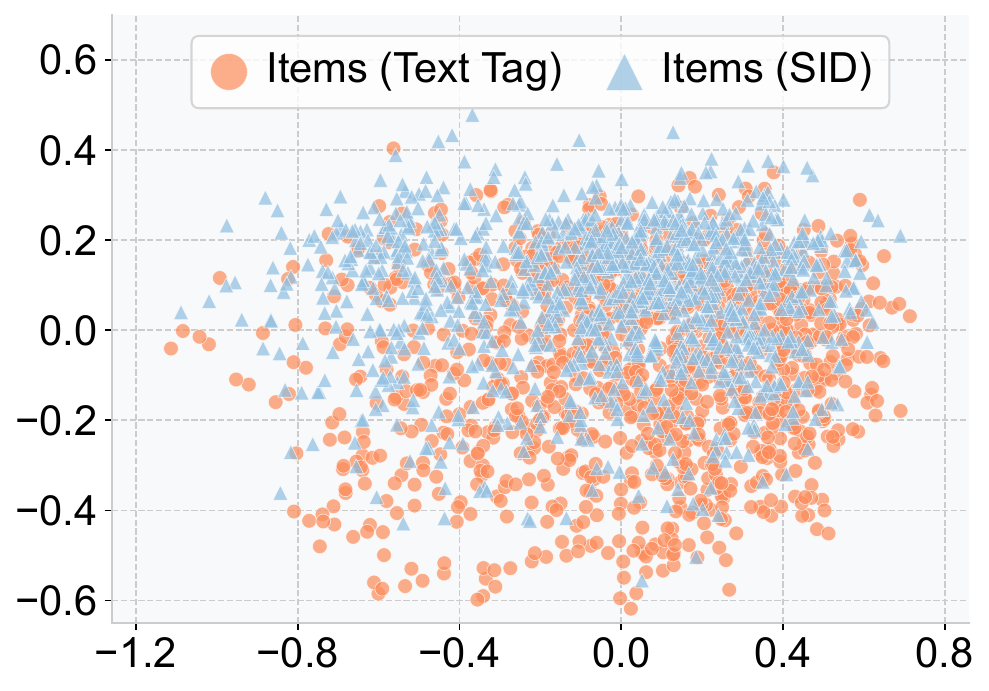}
\caption{PCA visualization of item embeddings retrieved by text tags and SIDs.}
\label{fig:pca-diversity}
\end{wrapfigure}
\paragraph{Embedding-Space Visualization.}
We further examine the diversity of items retrieved through text tags versus SIDs. For each text tag and its co-generated SIDs, we sample an equal number of relevant items and obtain their embeddings from a pre-trained retrieval model. As shown in Figure~\ref{fig:pca-diversity}, we project these embeddings to 2D via PCA. Tag-matched items show greater spatial dispersion across diverse regions, while SID-matched items form tighter clusters around specific neighborhoods. This pattern visually reinforces the world-knowledge-driven coverage of text tags versus the collaboration-driven specificity of SIDs.

\begin{table}[h]
\centering
\caption{Item-level hit rate under text tag, SID, and hybrid retrieval configurations.}
\label{tab:retrieval-hitrate}
\begin{tabular}{lcc}
\toprule
\textbf{Configuration} & \textbf{HR@500} & \textbf{HR@1000} \\
\midrule
Tag & 0.1503 & 0.2044 \\
SID & 0.1539 & 0.2144 \\
Hybrid & \textbf{0.1571} & \textbf{0.2168} \\
\bottomrule
\end{tabular}
\end{table}

\begin{figure*}[t]
  \centering
  \includegraphics[width=0.9\textwidth]{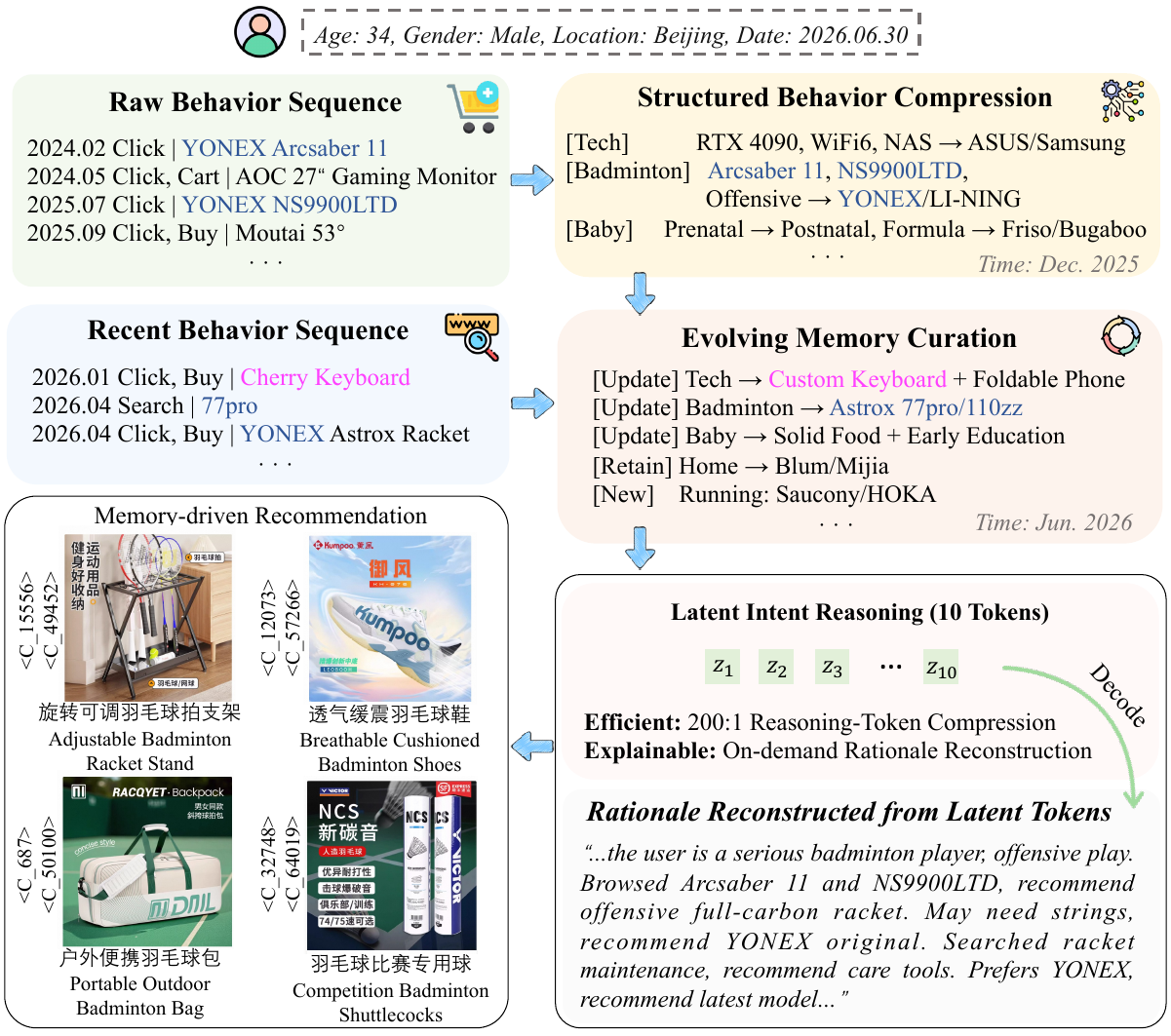}
  \caption{Case study. The memory hub compresses a user's raw behavior sequence into structured preference units and incrementally updates them with recent behaviors. Based on the curated memory, RecGPT-V3 performs latent intent reasoning with only 10 tokens, reconstructs an explainable rationale on demand, and generates memory-driven badminton recommendations grounded by SIDs.}
\label{fig:case_study}
\end{figure*}

\paragraph{Retrieval Complementarity.}
Finally, we quantify the end-to-end retrieval benefit of combining both modalities. Table~\ref{tab:retrieval-hitrate} reports item-level hit rate on an offline test set under three configurations. On HR@500, SID retrieval reaches 0.1539 compared to 0.1503 for text tag retrieval, and the hybrid configuration further improves to 0.1571. A consistent trend holds on HR@1000 (0.2168 vs.\ 0.2144 vs.\ 0.2044), confirming the complementarity of the two modalities.
The hybrid retrieval model (\S\ref{sec:retrieval}) consumes both modalities jointly, combining the broad coverage of text tags with the collaborative precision of SIDs.

\subsubsection{Case Study}
\label{sec:exp:case}

Figure~\ref{fig:case_study} presents an end-to-end case study of RecGPT-V3 on an anonymized Taobao user. The memory hub first compresses the raw behavior sequence into structured preference units, such as technology, badminton, and baby-care interests. When recent behaviors arrive, it selectively updates related memories: customized-keyboard interactions refresh the technology unit, Astrox-related searches and clicks update the badminton unit, baby-care interests evolve toward solid food and early education, stable home-improvement preferences are retained, and a new running preference is created.

Based on the curated memory, RecGPT-V3 performs Latent Intent Reasoning with only 10 latent tokens. These latent tokens replace the long explicit CoT during inference, but can still be decoded to reconstruct a readable explicit rationale on demand. In this case, the reconstructed rationale identifies the user as a serious badminton player with an offensive playing style and infers needs for full-carbon rackets, strings, maintenance tools, and related badminton equipment. These inferred intents are then grounded into concrete memory-driven recommendations with SIDs. This case shows how RecGPT-V3 accumulates user understanding through evolving memory, preserves explainable reasoning via latent-to-explicit CoT reconstruction, and generates precise recommendations from the inferred intent.
\section{Conclusion}
\label{sec:conclusion}

In this paper, we presented \recgpt[-V3], an LLM-based recommender that addresses three challenges facing LLM-based recommendation at industrial scale: stateless behavior modeling, the tag-to-item information bottleneck, and inefficient explicit reasoning. \recgpt[-V3] introduces a \emph{Memory Hub} that maintains a structured, continually evolving user memory in place of one-shot full-history encoding, a \emph{Hybrid-modal Foundation Model} that reasons jointly over natural language and Semantic IDs to ground intent directly in the item space, and \emph{Latent Intent Reasoning} that compresses explicit chain-of-thought into compact, decodable latent tokens. Deployed in Taobao's ``Guess What You Like'' feed, \recgpt[-V3] achieves consistent gains in large-scale online A/B tests (+1.28\% IPV, +1.00\% CTR, +1.97\% TC, and +3.97\% GMV) while reducing end-to-end serving compute by 52.4\%. These performance gains show that prediction accuracy and serving efficiency can improve together, and suggest that LLMs can serve as the reasoning brain of industrial-scale recommendation pipelines.

\addcontentsline{toc}{section}{References}
\bibliographystyle{abbrvnat}
\nobibliography*
\bibliography{reference}

\clearpage

\appendix
\section*{Appendix}
\definecolor{corecolor}{RGB}{34,102,68}
\definecolor{contribcolor}{RGB}{110,80,120}

\newcommand{\redact}{\rule[-0.15ex]{2.2em}{1.4ex}}

\section{Contributors}
\label{sec:contributors}

\noindent
\begin{minipage}[t]{0.48\textwidth}
\textbf{\textcolor{corecolor}{Core Contributors}}
\begin{itemize}[leftmargin=*, itemsep=1pt, topsep=4pt, label={}]
\color{corecolor!85!black}
    \item Bowen Zheng
    \item Chao Yi
    \item Dian Chen
    \item Gaoyang Guo
    \item Han Zhu
    \item Jiakai Tang
    \item Jian Wu
    \item Mao Zhang
    \item Wen Chen
    \item Yifan Lu
    \item Yujie Luo
    \item Yuning Jiang
    \item Zhujin Gao
\end{itemize}
\end{minipage}
\hfill
\begin{minipage}[t]{0.48\textwidth}
\textbf{\textcolor{contribcolor}{Contributors}}
\begin{itemize}[leftmargin=*, itemsep=1pt, topsep=4pt, label={}]
\color{contribcolor!85!black}
    \item Bo Zheng
    \item Chenchi Zhang
    \item Dixuan Wang
    \item Hao Fang
    \item Jiancai Liu
    \item Jing Yu
    \item Junjun Zheng
    \item Ke Chen
    \item Kewei Zhu
    \item Mengyan Li
    \item Mingke Xu
    \item Wenjun Yang
    \item Xiangheng Kong
    \item Xinming Zhang
    \item Xunke Xi
    \item Zile Zhou
\end{itemize}
\end{minipage}

\vspace{1em}
\noindent
The listing of authors is in alphabetical order based on their first names.

\section{Intent-to-Item Retrieval}
\label{sec:retrieval}

The Hybrid-modal Foundation Model (\S\ref{sec:foundation}) enables the multi-expert modules to express user intent through two channels: natural-language tags and Semantic IDs. The two play distinct roles: tags convey intent through open-ended world knowledge (\prop{generalizable}), whereas SIDs ground that intent in concrete item representations enriched by collaborative signals (\prop{concrete}). Turning these hybrid signals into accurate item retrieval, however, requires a dedicated downstream model that can (1)~consume both channels jointly, and (2)~balance click likelihood against semantic relevance. Earlier \recgpt{} versions retrieved items through a text-only pathway, confining the system to a single channel and leaving the collaborative signal carried by SIDs unexploited.

We therefore introduce a \textbf{hybrid retrieval model} that takes both text tags and SIDs as input and learns a preference ordering over candidate items, favoring those a user is likely to click while preserving semantic alignment with the upstream intent predictions. The remainder of this section presents its hybrid input architecture (\S\ref{sec:retrieval:input}) and joint training objective (\S\ref{sec:retrieval:objective}).

\subsection{Architecture}
\label{sec:retrieval:input}

The retriever takes the text tags and SIDs as input and assigns a relevance score to each candidate item. The two channels play distinct roles: a set of text tags $\{t_1, \ldots, t_M\}$ represents semantic intent, while a set of SIDs $\{d_1, \ldots, d_M\}$ grounds that intent in the item space. Each tag $t_i$ and its paired SID $d_i$ capture two sides of the same user intent: the former expresses ``\textcolor{observecolor!60!black}{what the user wants}'' in natural language, while the latter encodes ``\textcolor{hlcolor}{which items satisfy that want}'' in the discrete item space.

To fuse these two channels, we employ a dual-embedding architecture. Each tag $t_i$ is mapped to a dense vector via a text embedding layer $\mathbf{e}_t^{(i)} = \text{Emb}_{\text{tag}}(t_i)$, and each SID $d_i$ is independently mapped via a SID embedding layer $\mathbf{e}_d^{(i)} = \text{Emb}_{\text{sid}}(d_i)$. The two embeddings are concatenated and projected through a linear transformation to produce a unified intent representation:
\begin{equation}
    \mathbf{q}_i = W_{\text{proj}} \cdot [\mathbf{e}_t^{(i)} \| \mathbf{e}_d^{(i)}] + \mathbf{b}
    \label{eq:intent_fusion}
\end{equation}
where $[\cdot \| \cdot]$ denotes concatenation and $W_{\text{proj}} \in \mathbb{R}^{d \times 2d}$ is a learnable projection matrix. The resulting intent vectors $\{\mathbf{q}_1, \ldots, \mathbf{q}_M\}$ serve as queries in a Target-Attention mechanism, attending over the candidate item representations to compute their retrieval scores. This design preserves channel-specific information in the early layers while enabling cross-channel interaction through the shared attention computation.

The dual-channel design equips the retriever with collaborative evidence that a text-only model cannot access. Through the contrastive pre-training of \S\ref{sec:hybrid_tokenization}, the SID embedding space encodes item-level signals such as co-purchase patterns, popularity dynamics, and item-to-item affinity, which natural language cannot convey. The two channels also span different regions of the item space (\S\ref{sec:exp:sid}), and together they let the retriever pair the semantic breadth of language with the collaborative precision of discrete item representations.

\subsection{Training Recipe}
\label{sec:retrieval:objective}

Click signals alone provide insufficient supervision for the retriever. Because click propensity is confounded with item popularity, a purely click-driven objective tends to favor globally popular items over those faithful to the conditioning intent, producing retrievals that are clickable yet inconsistent with the tags and SIDs predicted upstream. This misalignment weakens the intent grounding that the pipeline is designed to preserve. We therefore cast retrieval training as a joint objective that reconciles click utility with semantic relevance under an explicit preference hierarchy over candidate items.

\paragraph{\textit{Preference Ordering.}}
We define a hierarchical preference over candidate items that integrates both utility and relevance signals:
\begin{tcolorbox}[colback=blue!5, colframe=white, arc=4pt, boxrule=0pt]
\centering
\textsc{Clicked} $\cap$ \textsc{Relevant} $\;>\;$ \textsc{Unclicked} $\cap$ \textsc{Relevant} $\;>\;$ \textsc{Unclicked} $\cap$ \textsc{Irrelevant}
\end{tcolorbox}
This ordering treats utility (click likelihood) as the primary objective and, among items of comparable utility, relies on relevance as the discriminating signal.

\paragraph{\textit{Joint Learning Objective.}}
The preference ordering translates into a joint training objective that couples two goals: a utility objective governing the primary click signal, and a relevance objective that, among items of comparable utility, resolves the ordering by semantic relevance. Optimizing the two jointly aligns click utility with intent grounding within a single objective.

In specific, the \textbf{utility loss} optimizes click prediction through a standard softmax objective over the item candidate set:
\begin{equation}
    \mathcal{L}_{\text{util}} = -\frac{1}{|\mathcal{C}|} \sum_{i \in \mathcal{C}} \log \frac{\exp(s_i)}{\exp(s_i) + \sum_{j \notin \mathcal{C}} \exp(s_j)}
    \label{eq:util_loss}
\end{equation}
where $\mathcal{C}$ denotes the set of clicked items and $s_i$ is the retrieval score for item $i$.

While the utility loss separates clicked items from the rest, it cannot order candidates that share the same click status. The \textbf{relevance loss} supplies this missing discrimination, enforcing the preference ordering through a multi-level contrastive formulation:
\begin{equation}
    \mathcal{L}_{\text{rel}} = -\frac{1}{N} \sum_{i=1}^{N} \frac{1}{|\mathcal{P}(i)|} \sum_{j \in \mathcal{P}(i)} \log \frac{\exp(s_j)}{\exp(s_j) + \sum_{k \in \mathcal{N}(i,j)} \exp(s_k)}
    \label{eq:rel_loss}
\end{equation}
where $\mathcal{P}(i)$ denotes the relevance-positive set for input intent $i$, constructed at progressively coarser matching criteria, and $\mathcal{N}(i,j)$ is the corresponding negative set. To keep the relevance objective from conflicting with the utility signal, we assign each candidate a priority by interaction depth, \emph{clicked} $>$ \emph{exposed} $>$ \emph{non-exposed}, and restrict $\mathcal{N}(i,j)$ to candidates of priority no higher than the positive $j$.

The final training objective is a weighted sum of the utility and relevance losses:
\begin{equation}
    \mathcal{L} = \alpha \, \mathcal{L}_{\text{util}} + \beta \, \mathcal{L}_{\text{rel}},
    \label{eq:retrieval_loss}
\end{equation}
where the coefficients $\alpha$ and $\beta$ balance the two objectives (we set $\alpha=1$ and $\beta=0.5$ in our production experiments). This joint objective drives the retriever to favor products that are click-worthy and semantically aligned with the upstream intent, coupling click utility with intent grounding within an end-to-end optimization framework.

\section{Latent Reasoning Reconstruction}
\label{app:reasoning_prompts}

The three alignment tasks of \S\ref{sec:reasoning:efficient} rely on a reconstruction prompt $\mathcal{P}$ that cues the model to decode the natural-language reasoning compressed into latent \texttt{<cot>} tokens.
We use two prompt variants, chosen by how many segments are being reconstructed: for \textbf{single- and multi-segment reconstruction} ($|\mathcal{J}| < K$), the prompt asks the model to explain the sentence each \texttt{<cot>} token likely represents; for \textbf{full-trace reconstruction} ($\mathcal{J} = \{1,\ldots,K\}$), it asks the model to recover the entire chain of thought.
Each reconstruction sample places the prompt in the system turn. The user turn then contains the original recommendation context (expert persona, user profile, and click history) followed by the \texttt{<think>} block, in which latent tokens replace the masked reasoning segments, and the model's final output. Given this input, the reconstruction model explains each masked \texttt{<cot>} token in natural language.
Below we present one representative production case per task. Personally identifiable details such as user location are redacted and shown as a blank bar (\redact).

\newcommand{\casesep}{\medskip\noindent\rule{\linewidth}{0.25pt}\medskip}
\definecolor{reconbg}{HTML}{E8F5E9}
\definecolor{reconframe}{HTML}{81C784}
\definecolor{scenariobg}{HTML}{E3F2FD}
\definecolor{scenarioframe}{HTML}{64B5F6}
\definecolor{sysbg}{HTML}{FFF3E0}

\clearpage
\setcounter{caseexample}{0}
\refstepcounter{caseexample}

\begin{tcolorbox}[enhanced,
  colback=white, colframe=gray!50,
  title={\textsf{Case~\thecaseexample: Single-Segment Reconstruction ($\mathcal{J}=\{1\}$)}},
  fonttitle=\bfseries\small,
  coltitle=white, colbacktitle=observecolor!65!black,
  arc=4pt, boxrule=0.5pt,
  top=4pt, bottom=4pt, left=5pt, right=5pt]

\scriptsize

\begin{tcolorbox}[enhanced, colback=white, colframe=gray!40,
  arc=3pt, boxrule=0.4pt,
  fonttitle=\bfseries\scriptsize, title={\textsf{Input}},
  top=0pt, bottom=0pt, left=0pt, right=0pt, middle=0pt]
\begin{tcolorbox}[colback=sysbg, colframe=sysbg, arc=0pt, boxrule=0pt,
  left=5pt, right=5pt, top=3pt, bottom=3pt]
\textbf{[System]}\; You are a semantic interpretation model skilled at explaining the meaning of \texttt{<cot>} tokens in a sentence. Output the sentences that the \texttt{<cot>} tokens in the following text likely represent, so that the resulting chain of thought is coherent.
\end{tcolorbox}
\begin{tcolorbox}[colback=scenariobg, colframe=scenariobg, arc=0pt, boxrule=0pt,
  left=5pt, right=5pt, top=3pt, bottom=3pt]
\textbf{[User]}\; You are a bags and accessories expert. User profile: \redact, residing in \redact, who needs to balance fashion and practicality in daily work and life, and prefers a comfortable, minimalist yet design-oriented style.
Interest pattern analysis:
- Repeatedly explores and purchases backpacks, soft-leather bags, and bags with rhinestone/diamond details; favors novel designs such as neo-Chinese vintage backpacks.
- Values both function and appearance: bags emphasize capacity (large capacity, dual-shoulder) and comfortable carrying, while also valuing durability features such as water resistance and abrasion resistance; likes visual accents such as rhinestones, color-blocking, and embroidery.
- Prefers practical bag styles suitable for switching across multiple occasions, ideal for urban commuting and short trips.
\ldots
Based on the user's historical preferences, please recommend women's bag products, and provide purchasing rationale considering design, practicality, and daily use. \ldots
\end{tcolorbox}
\begin{tcolorbox}[colback=thinkcolor!12, colframe=thinkcolor!12, arc=0pt, boxrule=0pt,
  left=5pt, right=5pt, top=3pt, bottom=3pt]
\texttt{<think>}\;\cottok{1}\\[2pt]
Possible recommended products: 1.~neo-Chinese vintage backpack (directly from click history)\quad 2.~rhinestone soft-leather backpack (combining preferences)\quad 3.~rhinestone embroidered crossbody bag \quad
4.~waterproof large-capacity backpack\quad
5.~color-blocked handbag\quad
6.~vintage-style shoulder bag\quad
7.~comfortable-carry backpack\quad
8.~rhinestone chain bag\quad
9.~neo-Chinese embroidered bag\quad
10.~soft-leather rhinestone wallet\\[2pt]
But we must ensure the tags are specific and relevant. From the click history, the user clicked ``neo-Chinese vintage backpack'', so similar styles should be recommended. \ldots\\
Ensure tags are specific: each tag has a core word such as ``backpack'' or ``crossbody bag'', with modifiers such as ``neo-Chinese'', ``rhinestone'', ``waterproof'', etc.\\
Finally, check relevance: all recommendations are based on user preferences and click history.\;\texttt{</think>}
\end{tcolorbox}
\begin{tcolorbox}[colback=gray!6, colframe=gray!6, arc=0pt, boxrule=0pt,
  left=5pt, right=5pt, top=3pt, bottom=3pt]
\texttt{[\{"t":~"neo-Chinese embroidered bag"\},~\{"t":~"soft-leather abrasion-resistant backpack"\}\ldots]}
\end{tcolorbox}
\end{tcolorbox}

\casesep

\begin{tcolorbox}[colback=reconbg, colframe=reconframe!70, arc=3pt, boxrule=0.3pt,
  left=4pt, right=4pt, top=3pt, bottom=3pt,
  fonttitle=\bfseries\scriptsize, title={\textsf{Reconstruction Output}}]
The meaning of the \cottok{1} token is as follows: \{\cottok{1}:
"First, the user is \redact, residing in \redact, who needs to balance fashion and practicality in daily work and life, and prefers a comfortable, minimalist yet design-oriented style. Her interest pattern includes: repeatedly exploring and purchasing backpacks, soft-leather bags, and bags with rhinestone/diamond details, favoring novel designs such as neo-Chinese vintage backpacks. \ldots\\
Task requirement: based on the above information, recommend the products the user is most likely to click next. Product tags must be specific and not too broad \ldots output at least 3 and at most 10.\\
Now, analyze the user's click history:\\
- Clicked products include: a waterproof sun-protective windbreaker, a stand-collar plus-size wool coat \ldots but also some bag-related ones, such as a waterproof large-capacity satchel, a handheld small square crossbody bag, an Oxford-cloth minimalist crossbody bag, a neo-Chinese vintage backpack, etc.\\
- Specific bag-related clicks: waterproof large-capacity satchel, handheld small square crossbody bag, Oxford-cloth minimalist crossbody bag, neo-Chinese vintage backpack. This indicates the user has a sustained interest in bags, especially backpacks and crossbody bags, with design elements such as neo-Chinese and rhinestones.\\
From the click history, the user's preferences are:\\
- Bag types: backpack, crossbody bag, handbag.\\
- Design elements: neo-Chinese, vintage, rhinestone, embroidery, color-blocking.\\
- Practical features: waterproof, large capacity, comfortable carrying.\\
- Materials: soft leather, genuine leather, sheepskin, etc.\\
\ldots\}
\end{tcolorbox}

\end{tcolorbox}

\clearpage
\refstepcounter{caseexample}

\begin{tcolorbox}[enhanced,
  colback=white, colframe=gray!50,
  title={\textsf{Case~\thecaseexample: Multi-Segment Reconstruction ($\mathcal{J}=\{1,2\}$)}},
  fonttitle=\bfseries\small,
  coltitle=white, colbacktitle=observecolor!65!black,
  arc=4pt, boxrule=0.5pt,
  top=4pt, bottom=4pt, left=5pt, right=5pt]

\scriptsize

\begin{tcolorbox}[enhanced, colback=white, colframe=gray!40,
  arc=3pt, boxrule=0.4pt,
  fonttitle=\bfseries\scriptsize, title={\textsf{Input}},
  top=0pt, bottom=0pt, left=0pt, right=0pt, middle=0pt]
\begin{tcolorbox}[colback=sysbg, colframe=sysbg, arc=0pt, boxrule=0pt,
  left=5pt, right=5pt, top=3pt, bottom=3pt]
\textbf{[System]}\; You are a semantic interpretation model skilled at explaining the meaning of \texttt{<cot>} tokens in a sentence. Output the sentences that the \texttt{<cot>} tokens in the following text likely represent, so that the resulting chain of thought is coherent.
\end{tcolorbox}
\begin{tcolorbox}[colback=scenariobg, colframe=scenariobg, arc=0pt, boxrule=0pt,
  left=5pt, right=5pt, top=3pt, bottom=3pt]
\textbf{[User]}\; You are a professional fashion apparel expert, particularly skilled at coordinating outfits for middle-aged women and teenagers.
User profile:
\redact~with certain requirements for apparel quality. \ldots

Apparel preferences analyzed from purchase history:
1. Favors a simple and elegant style: has purchased white shirts, knitwear, basic T-shirts
2. Values comfort: has purchased cotton and ice-silk garments
3. Has sun-protection needs: sun-protective clothing, long-sleeve shirts, etc.
4. Attends to the daughter's clothing: has purchased girls' clothing, school uniforms, etc.

Current environment and seasonal needs: \redact~summer is hot and humid (29--34\textdegree C), Major Heat is approaching, with light rain, and school starts in late July.
\ldots
Please recommend apparel and accessory products, including cool and breathable summer clothing, elegant pieces suitable for middle-aged women, sun-protective functional clothing, parent-child styles that can match the daughter, and semi-formal clothing suitable for the new school term. \ldots
\end{tcolorbox}
\begin{tcolorbox}[colback=thinkcolor!12, colframe=thinkcolor!12, arc=0pt, boxrule=0pt,
  left=5pt, right=5pt, top=3pt, bottom=3pt]
\texttt{<think>}\;\cottok{1}\;\cottok{2}\\[2pt]
- sun-protective functional clothing\quad - parent-child styles that can match the daughter\quad - semi-formal clothing suitable for the new school term\\[2pt]
Extracting trends from click history:\\
- Shirts: French-style blue shirt, French-style floral shirt~--~prefers French style and floral prints\\
- Dresses: printed sleeveless dress, striped waist-cinched dress, French-style gray dress~--~likes dresses with floral prints and waist-cinching elements\\
- Pants: Tencel-linen ice-silk Yamamoto pants, bohemian linen wide-leg pants, vintage floral wide-leg pants~--~likes loose, comfortable pants\\
- Shoes: mesh breathable sandals, mesh comfortable Mary Jane shoes, soft-sole non-slip nurse shoes~--~values comfort and breathability in shoes\\
\ldots\quad Recommended products should be based on relevance (French, vintage, minimalist), efficiency (urgently needed in summer), and diversity (covering tops, bottoms, dresses, shoes). \ldots\;\texttt{</think>}
\end{tcolorbox}
\begin{tcolorbox}[colback=gray!6, colframe=gray!6, arc=0pt, boxrule=0pt,
  left=5pt, right=5pt, top=3pt, bottom=3pt]
\texttt{[\{"t":~"French-style floral top"\},~\{"t":~"waist-cinched slimming dress"\}\ldots]}
\end{tcolorbox}
\end{tcolorbox}

\casesep

\begin{tcolorbox}[colback=reconbg, colframe=reconframe!70, arc=3pt, boxrule=0.3pt,
  left=4pt, right=4pt, top=3pt, bottom=3pt,
  fonttitle=\bfseries\scriptsize, title={\textsf{Reconstruction Output}}]
The meanings of the \cottok{1} and \cottok{2} tokens are as follows: \{\cottok{1}:
"First, the user is \redact, with requirements for apparel quality\ldots\\
Analysis from purchase history:\\
- Favors a simple and elegant style: white shirts, knitwear, basic T-shirts\\
- Values comfort: cotton and ice-silk materials\\
\ldots\\
Current environment: \redact~summer is hot and humid (29--34\textdegree C), the Major Heat solar term is approaching, so cool and breathable clothing is needed \ldots\\
\ldots\\
List of products the user clicked in the past month:\\
- French-style blue shirt\\
- bear-print Korean-style POLO shirt\\
\ldots\\
- soft-sole non-slip nurse shoes",\\[3pt]
\cottok{2}:
"\ldots\\
- blue waist-cinched suit\\
From the click history, the user clicked many apparel products, including shirts, dresses, pants, shoes, etc., with styles leaning toward French, vintage, and Korean, and materials such as Tencel, linen, and ice-silk, valuing comfort and breathability. \ldots\\
Recommended products need to be:\\
- cool and breathable summer clothing\\
- elegant pieces suitable for middle-aged women"\}
\end{tcolorbox}

\end{tcolorbox}

\clearpage
\refstepcounter{caseexample}

\begin{tcolorbox}[enhanced,
  colback=white, colframe=gray!50,
  title={\textsf{Case~\thecaseexample: Full-Trace Reconstruction ($\mathcal{J}=\{1,\ldots,8\}$)}},
  fonttitle=\bfseries\small,
  coltitle=white, colbacktitle=observecolor!65!black,
  arc=4pt, boxrule=0.5pt,
  top=4pt, bottom=4pt, left=5pt, right=5pt]

\scriptsize

\begin{tcolorbox}[enhanced, colback=white, colframe=gray!40,
  arc=3pt, boxrule=0.4pt,
  fonttitle=\bfseries\scriptsize, title={\textsf{Input}},
  top=0pt, bottom=0pt, left=0pt, right=0pt, middle=0pt]
\begin{tcolorbox}[colback=sysbg, colframe=sysbg, arc=0pt, boxrule=0pt,
  left=5pt, right=5pt, top=3pt, bottom=3pt]
\textbf{[System]}\; You are a semantic interpretation model skilled at recovering the chain of thought behind \texttt{<cot>} tokens in a sentence.
\end{tcolorbox}
\begin{tcolorbox}[colback=scenariobg, colframe=scenariobg, arc=0pt, boxrule=0pt,
  left=5pt, right=5pt, top=3pt, bottom=3pt]
\textbf{[User]}\; You are a professional automotive accessories and modification expert, familiar with all kinds of car parts and decorative products.
User profile: \redact, owns a \redact~car (imported), values quality of life.
Automotive purchase records: has purchased car stickers, in-car aromatherapy, windshield decals, wiper blades, etc., showing an interest in car decoration and maintenance.
Consumption traits: prefers cute cartoon-style decorations, values the combination of practicality and aesthetics.
Vehicle characteristics: the \redact~car is a compact luxury car, suitable for refined and practical modification plans.
\ldots
Please recommend products across dimensions such as interior decoration and ambiance creation, practical functional accessories, exterior body decoration, maintenance and care products, and safety and convenience equipment. \ldots
\end{tcolorbox}
\begin{tcolorbox}[colback=thinkcolor!12, colframe=thinkcolor!12, arc=0pt, boxrule=0pt,
  left=5pt, right=5pt, top=3pt, bottom=3pt]
\texttt{<think>}\;\cottok{1}\,\cottok{2}\,\cottok{3}\,\cottok{4}\,\cottok{5}\,\cottok{6}\,\cottok{7}\,\cottok{8}\;\texttt{</think>}
\end{tcolorbox}
\begin{tcolorbox}[colback=gray!6, colframe=gray!6, arc=0pt, boxrule=0pt,
  left=5pt, right=5pt, top=3pt, bottom=3pt]
\texttt{[\{"t":~"cute cartoon car ornament"\},~\{"t":~"sun-protective cartoon sunshade"\}\ldots]}
\end{tcolorbox}
\end{tcolorbox}

\casesep

\begin{tcolorbox}[colback=reconbg, colframe=reconframe!70, arc=3pt, boxrule=0.3pt,
  left=4pt, right=4pt, top=3pt, bottom=3pt,
  fonttitle=\bfseries\scriptsize, title={\textsf{Reconstruction Output}}]
First, the user is \redact, owns a \redact~ (imported) compact luxury car, and values quality of life. Purchase records show she has bought car stickers, in-car aromatherapy, windshield decals, wiper blades, etc.; she prefers cute cartoon-style decorations and values the combination of practicality and aesthetics. The vehicle is suitable for refined and practical modification plans.\\[2pt]
Demand dimensions:\\
1. interior decoration and ambiance creation\quad 2. practical functional accessories\quad 3. exterior body decoration\quad 4. maintenance and care products\quad 5. safety and convenience equipment\\[2pt]
Recommend 2--3 specific products per dimension, but the task requires adjusting the number of outputs based on the user's click history, at least 3 and at most 10. The list of products the user clicked in the past month shows she clicked many cute cartoon-style items, such as a polka-dot Miffy phone case, a new cartoon sunshade, a black-leather Kitty car ornament, a new Hello Kitty car ornament, etc. This indicates she has a strong preference for cute, cartoon-themed decorations.\\[2pt]
From the click history, the user clicked:\\
- many cartoon-themed items: Miffy, Kitty, bear, etc.\\
- car-related items: sunshade, car ornament, car mat, etc.\\
\ldots\\
Therefore, the recommended products should be car-related and incorporate a cute cartoon style, while also considering practicality and aesthetics.\\[2pt]
Now, analyze each dimension and recommend specific products:\\
1. \textbf{Interior decoration and ambiance creation:}\ldots Specifically: cute cartoon car ornament, aromatherapy air-vent clip, ambient lighting.\\
2. \textbf{Practical functional accessories:}the user clicked sunshades, car mats, etc. Recommend: sun-protective sunshade curtain, comfortable seat cushion, storage box.\\
3. \textbf{Exterior body decoration:}the user has purchased car stickers and windshield decals. Recommend: cartoon-pattern car stickers \ldots\\
\ldots\\
All recommendations incorporate a cute cartoon style, matching the user's aesthetic preferences, ensuring tags are specific and relevant.
\end{tcolorbox}

\end{tcolorbox}

\end{CJK*}
\end{document}